\documentclass[conference]{IEEEtran}
\usepackage{cite}
\usepackage{amsmath,amssymb,amsfonts}
\usepackage{algorithmic}
\usepackage{graphicx}
\usepackage{textcomp}
\usepackage{xcolor}
\usepackage[hyphens]{url}
\usepackage{fancyhdr}
\usepackage[bookmarks=true,breaklinks=true,letterpaper=true,colorlinks,citecolor=blue,linkcolor=blue,urlcolor=blue]{hyperref}
\usepackage{listings}

 \usepackage{bbding}
\author{
{Zhe Zhou\textsuperscript{*1,2}, Yiqi Chen\textsuperscript{*1}, Tao Zhang\textsuperscript{4}, Yang Wang\textsuperscript{4}, Ran Shu\textsuperscript{4}, Shuotao Xu\textsuperscript{4},  Peng Cheng\textsuperscript{4}, Lei Qu\textsuperscript{4}} \\ {Yongqiang Xiong\textsuperscript{4},   Jie Zhang\textsuperscript{2,5},  Guangyu Sun\textsuperscript{$\dagger$}\textsuperscript{1,3}} \vspace{0.5em} \\
\textsuperscript{1}\emph{School of Integrated Circuits}, \textsuperscript{2}\emph{School of Computer Science}, 
\emph{Peking University}\\
\textsuperscript{3} \emph{Beijing Advanced Innovation Center for Integrated Circuits}\\
\textsuperscript{4} \emph{Microsoft Research},
\textsuperscript{5} \emph{Zhongguancun Laboratory} \\
\emph{\{zhou.zhe, yiqi.chen, jiez, gsun\}@pku.edu.cn} \\  \emph{\{zhangt, yang.wang, ran.shu, shuotaoxu, pengc, lei.qu, yongqiang.xiong\}@microsoft.com}\\
}

\title{NeoMem: Hardware/Software Co-Design for CXL-Native Memory Tiering} 

\newcommand{\rev}[1]{{#1}}
\usepackage{makecell}
\usepackage{multirow} 
\usepackage[normalem]{ulem}
\usepackage[linesnumbered, ruled, vlined]{algorithm2e}
\usepackage{pifont}
\usepackage{amsfonts}
\usepackage{framed}
\usepackage{graphicx}
\newcommand{\bone}{\ding{182}}
\newcommand{\btwo}{\ding{183}}
\newcommand{\bthree}{\ding{184}}
\newcommand{\bfour}{\ding{185}}
\newcommand{\bfive}{\ding{186}}
\newcommand{\bsix}{\ding{187}}
\newcommand{\bseven}{\ding{188}}
\newcommand{\beight}{\ding{189}}

\usepackage{booktabs}
\usepackage{hyperref} 
\hypersetup{hypertex=true,
colorlinks=true,
linkcolor=black,
anchorcolor=black,
citecolor=black}

\usepackage{titlesec}
\makeatletter
\g@addto@macro{\normalsize}{%
  \setlength{\abovedisplayskip}{3pt plus 0.5pt minus 1pt}
  \setlength{\belowdisplayskip}{3pt plus 0.5pt minus 1pt}
  \setlength{\abovedisplayshortskip}{0pt}
  \setlength{\belowdisplayshortskip}{0pt}
  \setlength{\intextsep}{4pt plus 1pt minus 1pt}
  \setlength{\textfloatsep}{4pt plus 1pt minus 1pt}
  \setlength{\skip\footins}{5pt plus 1pt minus 1pt}
  }
\makeatother

\titlespacing\section{0pt}{2pt plus 1pt minus 1pt}{3pt plus 1pt minus 2pt}
\titlespacing\subsection{0pt}{2pt plus 1pt minus 1pt}{3pt plus 1pt minus 2pt}
\titlespacing\subsubsection{0pt}{2pt plus 1pt minus 1pt}{3pt plus 1pt minus 2pt}

\usepackage{enumitem}

\lstdefinestyle{Python}{
	language        =   bash, 
	keywordstyle    =   \color{blue},
	keywordstyle    =   [2] \color{teal},
	stringstyle     =   \color{magenta},
	commentstyle    =   \color{red}\ttfamily,
	breaklines      =   true,   
	columns         =   fixed,  
	basewidth       =   0.5em,
}

\begin{document}
\maketitle
\pagestyle{plain}

\begin{abstract}

The Compute Express Link (CXL) interconnect  makes it feasible to integrate diverse types of memory into servers via its byte-addressable SerDes links.  Considering the various access latency, harnessing the full potential of CXL-based heterogeneous memory systems requires efficient memory tiering. However,  prior work can hardly make a fundamental progress owing to  low-resolution and high-overhead memory access profiling techniques.
To address this critical challenge, we propose  a novel memory tiering solution called  \emph{NeoMem}, which features a hardware/software co-design. 
NeoMem 
offloads memory profiling functions to CXL device-side controllers, integrating a dedicated hardware unit called \emph{NeoProf}. NeoProf readily monitors memory accesses and provides the OS with crucial  page hotness statistics and other useful system state information. 
On the OS kernel side, we design a revamped memory-tiering strategy, enabling accurate and timely hot page promotion based on NeoProf statistics. 
We implement NeoMem on a real  FPGA-based CXL memory platform and Linux kernel v6.3.  Comprehensive evaluations demonstrate that NeoMem achieves {$32\% \sim 67\%$} geomean speedup over several existing memory tiering solutions.

\newcommand\blfootnote[1]{%
\begingroup
\renewcommand\thefootnote{}\footnote{#1}%
\addtocounter{footnote}{-1}%
\endgroup
}

\blfootnote{* Co-first authors.}
\blfootnote{$\dagger$ Corresponding author.}

\end{abstract}
\section{Introduction}

\label{sec:intro}

\begin{figure} [t]
    \centering
    \includegraphics[width=0.98\linewidth]{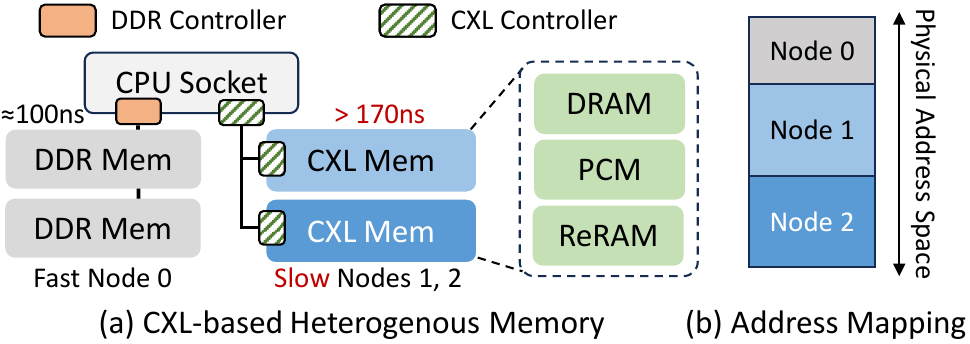} 
    \vspace{-0.5em}
    \caption{A Typical CXL-based Tiered Memory System.}
         \label{fig:cxl_mem}
\end{figure}

The Compute Express Link (CXL) technology  provides a coherent and byte-addressable interconnect between host CPUs and various peripheral devices~\cite{cxl}. Among its multiple use cases~\cite{cxl-anns,  sim2022computational, kwon2023cache, arif2023accelerating, zhang2023rethinking, cho2023case, banakar2023wiscsort}, CXL-based memory extension (CXL memory for short) has risen as a focal point~\cite{tpp,pond,ha2023dynamic,gouk2022direct, boles2023cxl}. As  illustrated in Figure~\ref{fig:cxl_mem}-(a), CXL enables servers  to  incorporate diverse memory devices to expand the memory capacity and bandwidth without necessitating hardware modifications on the CPU side.  As depicted in Figure~\ref{fig:cxl_mem}-(b),  CXL memories are usually exposed to the OS as CPU-less NUMA nodes~\cite{tpp, pond}. CPUs  directly access these address-mapped NUMA nodes without invoking page faults or swaps. 

However, CXL memory exhibits the elevated access latency, which can be over twice as high as that of DDR DRAMs~\cite{intel-fpga, SMT, pond}. 
The latency is more pronounced when replacing DRAM with slower memory media like PCM and ReRAM in the CXL memory devices~\cite{sha2023vtmm}. 
Therefore, in a system with local DDR DRAM and diverse CXL memories, the disparities in memory latency, bandwidth and capacity result in the formation of a  tiered memory system~\cite{sha2023vtmm, pond, tpp, lee2023memtis}.
Typically, accesses to faster memory tiers are characterized by lower  latency and higher bandwidth, whereas the opposite holds true for slower tiers. Given this differential, the OS should place frequently accessed ``hot" pages in fast memory tiers while putting ``cold" pages in slow tiers to maximize system performance, which is referred to as \emph{Memory Tiering} technique~\cite{raybuck2021hemem, yan2019nimble, agarwal2017thermostat, heteroOS,autotiering, tmts}. Obviously,  efficient hot page detection methods are crucial to memory tiering. 

However, realizing efficient hot page detection in CXL-based tiered memory systems is challenging, mainly due to the low-resolution and high-overhead memory access profiling techniques.
Unlike RDMA-based memory disaggregation, where the OS monitors external memory access~\cite{al2020effectively,gu2017efficient}, CPU directly accesses CXL memory via traditional load/store instructions without OS awareness of the access patterns. Therefore, existing memory tiering techiniques rely on special profiling methods like PTE-scan~\cite{damon}, hint-fault monitoring~\cite{gandhi2014badgertrap}, and PMU sampling~\cite{weaver2016advanced}, each with inherent limitations, to gain visibility to memory access.   

To be specific, 
PTE-scan periodically clears \emph{Access} bits in Page-Table Entries (PTEs). It then identifies accessed pages by scanning the page table.  PTE-scan 
only captures one access per page in each scanning epoch, leading to low resolution.   In hint-fault monitoring~\cite{gandhi2014badgertrap}, the OS ``poisons" some  PTEs by setting special bits.  The following accesses to these  pages will trigger protection faults, immediately notifying the OS with the page access. However, due to frequently triggering page faults,  hint-fault monitoring  incurs severer overhead~\cite{ren2023hm}. 
Moreover, both techniques operate at the TLB level and cannot detect  true LLC misses. {If a page is frequently accessed but always resides in  cache, it is redundant to migrate it to fast memory. }


Performance Monitoring Units (PMUs) in CPUs, such as Intel's PEBS and AMD's IBS,  can be employed to track LLC misses via sampling events. 
While PMU-sampling directly tracks LLC misses, it always operates at a low sampling frequency to control overhead~\cite{tmts, lee2023memtis}, hindering the achievement of optimal hot page detection recall in practical scenarios~\cite{tmts}.  Additionally,  {PMU-based methods are CPU vendor-specific, which limits their generality.} A detailed analysis of these memory profiling techniques is presented in Section~\ref{sec:memory_profiling}.

Given these challenges, Linux developers have reached a consensus~\cite{future} that \emph{``The biggest problem for memory tiering still appears to be page promotion"} because  \emph{``It is  difficult to determine when a page has become hot"}. It is urgent to devise efficient and practical memory access profiling and hot page detection techniques for  CXL-based memory tiering. 

\begin{figure} [t]
    \centering
    \includegraphics[width=0.98\linewidth]{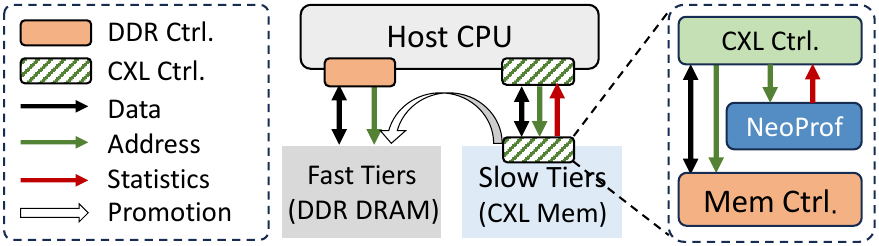} 
    \vspace{-0.5em}
    \caption{Illustration of NeoProf-based Memory Profiling}
         \label{fig:neomem_abstract}
\end{figure}

\noindent\textbf{Design goals:} 
We claim that an ideal memory access profiling mechanism should aim to fulfill the following goals.

\noindent\emph{\textbf{G1}: {High Resolution} }– First of all, the profiler should accurately and promptly track the frequency and location of memory accesses with high time  and space resolution.

\noindent\emph{\textbf{G2}: {Low Overhead}} – It is also crucial for the profiler to consume minimal CPU cycles, ensuring that system's performance is not detrimentally impacted.

\noindent\emph{\textbf{G3}: {Cache Awareness}} – The profiler should be aware of CPU's cache and only captures LLC misses to ensure a high accuracy.

\noindent\emph{\textbf{G4}: {Universal Compatibility}} – The design should be versatile, ensuring compatibility across various platforms, provided they support CXL-based memory expansion.

\noindent\emph{\textbf{G5}: {Comprehensive Profiling}} – 
Additionally, if the profiler can also capture crucial runtime information such as system bandwidth utilization, read/write distribution and access frequency distribution, etc., the OS can perform a better scheduling by leveraging these information.

\noindent\textbf{Our Work:}
We capitalize on the special architecture of CXL memory and propose \emph{NeoMem}, a \emph{CXL-native}\footnote{We call NeoMem ``CXL-native" because it fully utilizes the special architecture of CXL memory that has device-side controllers.} memory tiering solution featuring a {hardware-software} co-design. NeoMem integrates memory access profiling units, named \emph{NeoProf}, into CXL memories' device-side controllers, as illustrated in Figure~\ref{fig:neomem_abstract}. NeoProf readily analyzes LLC misses to CXL memory and provides the OS with crucial information like page hotness, memory bandwidth utilization, read/write ratio, and access frequency distribution, etc. Meanwhile,  on the OS side, we design an advanced memory-tiering strategy by leveraging NeoProf's insights for efficient hot page promotion. We demonstrate how  NeoMem meets the five design goals: 

For  goal \textbf{G1},  NeoProf uses a customized Sketch-based~\cite{cmsketch} hot-page detector to efficiently analyze \emph{each} physical page access and identify hot pages with a fine granularity of 4KB. 
By offloading hot page detection to dedicated hardware, NeoMem saves precious CPU cycles, which  inherently satisfies \textbf{G2}. As NeoProf resides at the CXL memory side, it  directly monitors true LLC misses,  which naturally fulfills~\textbf{G3}. 
To ensure broad compatibility, as outlined in \textbf{G4}, we limit hardware modifications, namely NeoProf, to the CXL device side. This guarantees seamless integration with any host CPUs. Beyond tracking page hotness, NeoProf offers insights into other vital metrics, such as bandwidth utilization, read/write ratio and access frequency distribution, etc. These statistics empower the OS to dynamically control page migration aggressiveness for optimal performance, which fulfills \textbf{G5}.

In contrast to previous studies using  emulation to prototype their designs~\cite{pond, tpp, sha2023vtmm}, NeoMem is validated on a \textbf{real}  FPGA-based CXL memory platform. 
We implement NeoMem's driver and memory tiering daemon in the Linux kernel v6.3. Both the software and hardware components are publicly available in the provided repository\footnote{~\url{https://github.com/PKUZHOU/NeoMem-MICRO-2024}}.
To summarize, we have  made the following key contributions: 
\begin{itemize}[leftmargin=1em]
\item \emph{Limitation Analysis.} We investigate the limitations of existing memory tiering methods and demonstrate the necessity of an efficient memory access profiling technique in emerging CXL-based tiered memory systems (Sec. \ref{sec:background}). 
\item \emph{{NeoMem Solution}.}  We propose a novel NeoMem {solution}, which leverages a dedicated hardware profiler, NeoProf, in memory-side controller to realize efficient memory profiling (Sec. \ref{sec:neomem}).  We carefully design the  architecture of NeoProf to ensure high profiling accuracy and low overhead (Sec.~\ref{sec:neoprof}). Based on NeoProf, we introduce NeoMem's software design and dynamic migration policy (Sec. \ref{sec:neomem_software}).
\item \emph{Real-Platform Prototyping.} We conduct real-platform prototyping of NeoMem based on a CXL-enabled FPGA platform and Linux kernel v6.3 with our NeoMem patch (Sec. \ref{sec:eval}). 
\end{itemize}

 According to our evaluation, NeoMem achieves {$32\%$ to $67\%$} geomeam speedup on eight representative benchmarks compared to several existing memory-tiering solutions.  
\noindent 
\section{Background and Motivation}
\label{sec:background}

\begin{figure} [t]
    \centering
    \includegraphics[width=0.98\linewidth]{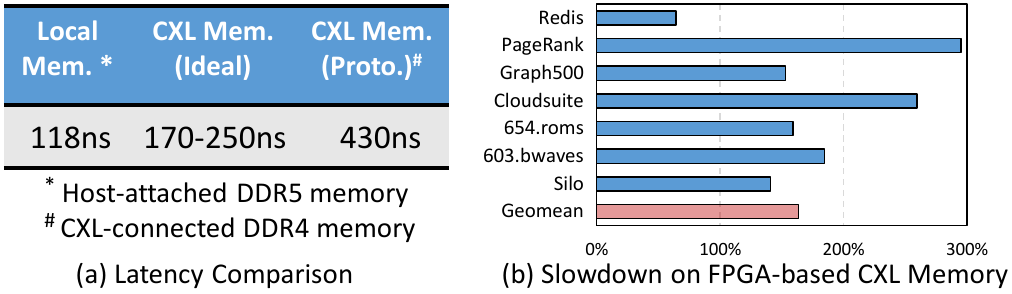} 
    \vspace{-0.5em}
    \caption{{Characterizing CXL-enabled Commodity Hardware.}}
         \label{fig:bg_slowdown}
\end{figure}

\begin{figure*} [t]
    \centering
    \includegraphics[width=1.0\linewidth]{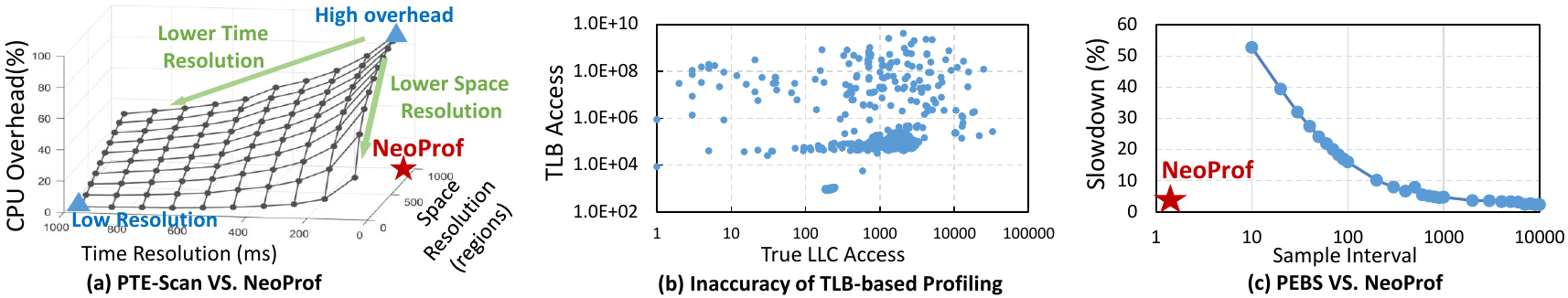} 
    \vspace{-1.2em}
    \caption{Evaluating Different Memory Profiling Mechanisms.}
         \label{fig:motivations}
         \vspace{-1em}
\end{figure*}

\subsection{CXL-based Tiered Memory System}

\label{sec:cxl_tiered_memory}


CXL (Compute-Express-Link)~\cite{cxl}, built on the PCIe 5.0 physical layer, creates a cache-coherent, byte-addressable interconnect through efficient SerDes links. It comprises three sub-protocols: \texttt{CXL.io}, \texttt{CXL.cache}, and \texttt{CXL.mem}. The \texttt{CXL.mem} protocol specifically allows  CPU to directly access CXL memory devices with load/store instructions. Unlike traditional DDR-based memory, where the memory controller is embedded in the host CPUs and has challenges to support new types of memories, CXL memory employs device-side  controllers and interact with host CPU via standard CXL protocol. This decoupled architecture facilitates the integration of various  types of memory  into servers, catering to specific capacity, performance, and cost needs~\cite{hynix_cxl_mem,samsung_cxl_mem, pond,tpp}.

However, compared to CPU-attached DDR-DRAMs, CXL-connected memories are considered to exhibit higher latency. On the one hand, the control and transmission overhead of CXL is non-negligible\cite{tpp,pond, intel-fpga, SMT}. On the other hand,  the latency is more pronounced when integrating high-density but slower memory media like PCM and ReRAM as CXL memory~\cite{sha2023vtmm, cxl-anns}.   
To delve deeper into this difference, we conduct a performance characterization on {Intel's FPGA-based} CXL memory prototype~\cite{sun2023demystifying}, with its detailed configurations provided in Section~\ref{sec:exp_setup}.


As shown in Figure~\ref{fig:bg_slowdown}-(a), Intel's CXL memory prototype exhibits a latency of approximately 430 ns, around $3.6\times$ that of the faster CPU-attached DDR-DRAM. Prior studies~\cite{tpp, pond}  tend to assume a CXL-memory latency of 170-250ns through NUMA-based emulation, less than the available prototype but still up to twice as high compared to fast-tier DRAM memory. Considering that this paper's goal is to propose  efficient memory access profiling and hot page detection mechanisms, rather than reduce CXL memories' latency, {the sub-optimal latency of Intel's prototype will not affect our conclusions.}

We also compare the end-to-end performance using several benchmarks (introduced in Section~\ref{sec:exp_setup}). We bound CPU threads to either the fast or slow memory tier to assess the resultant slowdown. As indicated in Figure~\ref{fig:bg_slowdown}-(b), solely utilizing the CXL memory results in a performance decrease ranging from $64\%$ (in Redis) to $295\%$ (in Page-Rank). These observations align with prior findings~\cite{intel-fpga}.

\subsection{Tiered Memory Management}
\label{sec:tmm}
In a system with heterogeneous types of memories, 
memory tiering techniques are usually adopted to maximize system performance. Memory tiering fundamentally hinges on the   data locality of workloads and  strategically places the frequently accessed ``hot" pages, which are performance critical, in faster memory tiers. 
The iterative memory tiering process typically unfolds in three key stages:

\begin{itemize}[leftmargin=1em]
    \item \emph{Memory-Access Profiling}. The system first gathers  page access statistics via specific memory profiling techniques.
    \item \emph{Page Classification}. Profiled pages are then classified as either ``hot" or ``cold" based on their access frequency.
    \item \emph{Page Migration.} A process, termed as \emph{Promotion}, migrates the hot pages identified in slow memory tiers to fast tiers. Conversely, a process, known as \emph{Demotion}, relocates the cold pages to slower memory tiers.  
\end{itemize}

\noindent  Among the three steps,  
memory-access profiling plays a fundamental role~\cite{damon,choi2021dancing,  ren2023rethinking,   raybuck2021hemem, sha2023vtmm}. A high-resolution and low-overhead memory profiling method is critical for promptly identifying performance-critical ``hot" pages. 
Since physical memory accesses are not visible to the operating system, several special profiling techniques have been proposed, which are analysed as follows.

\vspace{0.2em}
\subsection{Memory Access Profiling Techniques} 
\label{sec:memory_profiling}

\noindent\textbf{PTE-scan.} Page-Table-Entry scanning (PTE-scan) is a widely used method for page access tracking~\cite{heteroOS,yan2019nimble,amp,tmts, agarwal2017thermostat}. In this approach, a daemon thread in the OS kernel periodically resets the \emph{Accessed} bits in PTEs. When a page is accessed by the CPU, the corresponding bit in the PTE is set. { After a certain interval after previous resets, the daemon thread  scans the PTEs  to check which pages have been accessed. Therefore, PTE-scan tracks memory access in distinct epochs.} 

However,  PTE-scan faces challenges with low time resolution and high overhead.   
For instance, scanning PTEs in a large-scale memory system can even take several seconds~\cite{raybuck2021hemem}. Furthermore, PTE-scan  can only detect a single access per page in each scanning epoch, necessitating multiple epochs to identify frequently accessed pages, which  escalates costs and reduces timeliness.
Efforts like DAMON~\cite{damon} and AMP~\cite{amp} employ region sampling or huge pages to mitigate PTE scanning overhead. They however compromise on space resolution.

 We analyze the trade-offs in PTE-scan using  DAMON~\cite{damon} in Linux.  DAMON allows customization of both time resolution (interval between scans, in milliseconds) and space resolution (number of monitored regions, where a higher number equates to finer granularity). As depicted in Figure~\ref{fig:motivations}-(a), our analysis reveals that to achieve prompt memory access tracking with manageable overhead in PTE-scan, there's a significant compromise in space resolution, and vise-versa. 

\vspace{5pt}
\noindent\fbox{%
  \parbox{0.47\textwidth}{%
     \textbf{{Challenge\#1:}} PTE-scan methods cannot achieve high time and space resolution while maintaining a low overhead. 
  }%
}
\vspace{5pt}

\noindent\textbf{Hint-fault Monitoring.} 
In contrast to PTE-scan, which gathers page-access information in distinct epochs,  hint-fault monitoring can achieve more immediate page-access tracking~\cite{tpp, numabalancing, autotiering, agarwal2017thermostat,bergman2022reconsidering}. For example, Thermostat~\cite{agarwal2017thermostat} periodically samples  a subset of pages  and ``poisons" the corresponding PTEs by setting protection bits. Successive page accesses to these poisoned pages invokes protection faults immediately, signaling to the OS which pages have been accessed. 
However, as each page tracking operation initiates a costly TLB shootdown and page fault, hint-fault monitoring necessitates a sampling approach to temper overheads, which results in a low coverage in practice~\cite{ren2023rethinking}.

It is worth noting that both PTE-scan and hint-fault monitoring  track TLB misses  rather than LLC misses (a.k.a, true CXL memory access).  
Our detailed profiling reveals that TLB misses and LLC misses may not exhibit a strong correlation across various workload traces. This observation is illustrated in Figure~\ref{fig:motivations}-(b), where we graph the total number of TLB accesses (plotted on the Y-axis) against the LLC accesses (on the X-axis) for  sampled pages from a Redis~\cite{redis} trace. 
We utilize the KCacheSim~\cite{calciu2021rethinking} simulator for this analysis. {The memory accesses are filtered by 32KB/core L1D/I caches and 2MB/core L2 cache.}
The scatter plot in the figure clearly demonstrates a high level of dispersion, indicating that a page with frequent TLB accesses does not necessarily have a high number of LLC misses. 

\vspace{3pt}
\noindent\fbox{%
  \parbox{0.47\textwidth}{%
     \textbf{{Challenge\#2:}} TLB-based memory profiling methods may not capture the actual CXL memory access situation. 
  }%
}
\vspace{3pt}
\begin{table}[t]
\caption{Memory-Access Profiling Techniques Comparison}
\vspace{-0.5em}
\label{tab:prof_methods}
\centering
\setlength{\tabcolsep}{0.9mm}{
\resizebox{0.5\textwidth}{!}{\renewcommand{\arraystretch}{0.6}{
\begin{tabular}{c|c|c|c|c}
\toprule
                                                               & PTE-Scan                                                            & \begin{tabular}[c]{@{}c@{}}Hint-fault \\ Monitoring\end{tabular}       & \begin{tabular}[c]{@{}c@{}}PMU \\ Sampling\end{tabular} & \textbf{NeoProf}                                                               \\ \midrule
\begin{tabular}[c]{@{}c@{}}Profiling\\ Location\end{tabular}   & TLB                                                                 & TLB                                                                    & PMU Monitor                                             & \begin{tabular}[c]{@{}c@{}}\textbf{Device-side} \\ \textbf{CXL Controller}\end{tabular} \\ \midrule
\begin{tabular}[c]{@{}c@{}}Profiling\\ Resolution\end{tabular} & \begin{tabular}[c]{@{}c@{}}One Access \\ Per Epoch\end{tabular} & \begin{tabular}[c]{@{}c@{}}One Access to \\ Sampled Pages\end{tabular} & \begin{tabular}[c]{@{}c@{}}Sampled \\ Accesses\end{tabular}                                         & \textbf{Each Access}                                                           \\ \midrule
\begin{tabular}[c]{@{}c@{}}Cache\\ Aware?\end{tabular}         & \textcolor{red}{\XSolidBrush}                                                                   & \textcolor{red}{\XSolidBrush}                                                                       & \textcolor{green}{\CheckmarkBold}                                                       & \textcolor{green}{\CheckmarkBold}                                                                     \\ \midrule
Overhead                                                       & High                                                                & High                                                                   & Medium                                                  & \textbf{Low}                                                                   \\ \bottomrule
\end{tabular}
}

}}
\end{table}

\noindent\textbf{PMU Sampling.}
PMU (Performance Monitoring Unit) sampling
utilizes hardware monitors in  CPUs  for direct LLC miss tracking. Intel's PEBS, for instance, supports sampling LLC misses and storing them in a dedicated memory buffer. When this buffer reaches capacity, it triggers an interrupt, prompting the kernel to process these samples.

This direct tracking method is prevalent in memory tiering systems~\cite{raybuck2021hemem, lee2023memtis, tmts}.  However, as illustrated in Figure~\ref{fig:motivations}-(c), the overhead  of PEBS increases  with sampling frequency. Decreasing the sampling interval from every 10,000 to every 10 LLC misses can cause a workload slowdown of more than 50\%. Consequently, existing systems usually adopt a low sampling frequency to minimize performance impact, which however scarifies resolution~\cite{tmts, lee2023memtis}. 
Furthermore, PMU sampling is often closely tied to specific CPU platforms. Consequently, a system designed for Intel CPU servers may not seamlessly transition to AMD or ARM servers, which contradicts the open and versatile principles of CXL. 

\vspace{3pt}
\noindent\fbox{%
  \parbox{0.47\textwidth}{%
     \textbf{{Challenge\#3:}} The  PMU-sampling methods are CPU-vendor specific and can hardly achieve a high profiling resolution. 
  }%
}
\vspace{3pt}

\section{NeoMem {Solution}}
\label{sec:neomem}

To overcome these limitations, we propose a novel \emph{CXL-native}  memory tiering solution named NeoMem. 
NeoMem's key design philosophy is to offload the  memory profiling functions from CPU to  a dedicated hardware unit in CXL memory's controller, which is named NeoProf in our design. NeoProf readily monitors \emph{each} memory access to CXL memory in a page granularity and direcly provides the OS with crucial page hotness statistics and other useful runtime state information, avoding wasting precious CPU cycles for profiling. Based on NeoProf's statistics, the OS  performs accurate and timely hot page promotion with a revamped memory-tiering strategy. 
As depicted in  Figure~\ref{fig:motivations}-(a) and Figure~\ref{fig:motivations}-(c) and compared in Table~\ref{tab:prof_methods},  NeoProf achieves exceptionally high profiling resolution while imposing minimal CPU overhead.



\begin{figure} [t]
    \centering
    \includegraphics[width=0.98\linewidth]{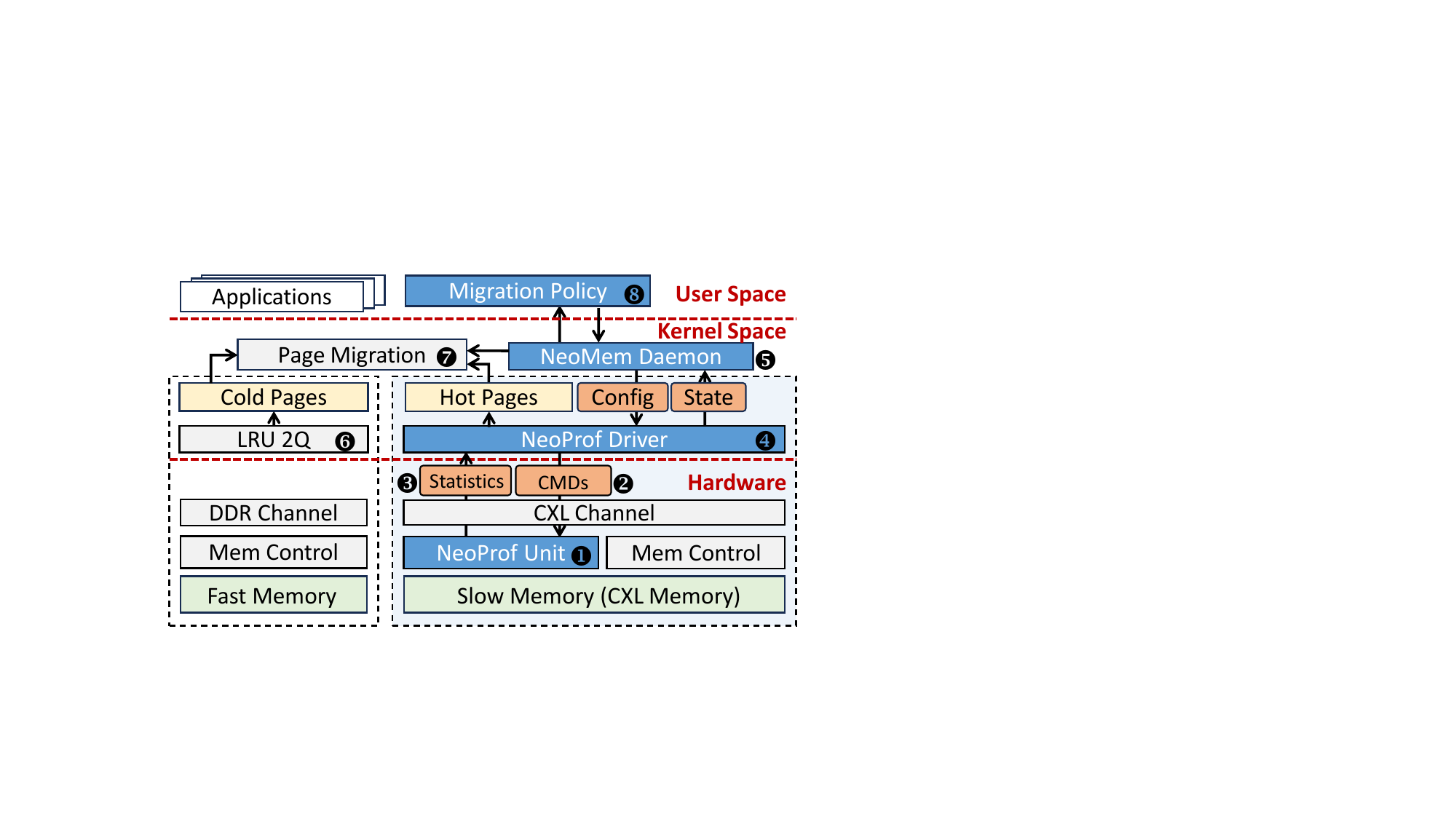} 
    \caption{Overview {of NeoMem Solution}.}
         \label{fig:system_stack}
\end{figure}

\subsection{{Overview of NeoMem Solution}}
\label{sec:stack_overview}
Figure~\ref{fig:system_stack} presents the  overview of NeoMem. In the  system, we currently assume two memory tiers:  the CPU-attached DDR DRAM serves as  the fast memory tier, while CXL-connected memory is the slow memory tier. Different memory tiers are managed via the NUMA APIs of Linux. NeoMem  introduces the following key components:

\noindent\textbf{Memory Access Profiling.} A pivotal feature of NeoMem is its efficient memory access profiling capability. NeoMem facilitates two core profiling functions: (1)  hot page detection in CXL memory and (2) monitoring of runtime states.

\noindent$\bullet$ \emph{Hot Page Detection.}
As shown in Figure~\ref{fig:system_stack}, NeoProf (\bone)  resides in CXL memory's device-side controllers. It snoops memory-access requests sent via the CXL channel,  analyzes them and generates page hotness information as well as other useful statistics (\bthree). The host CPU controls NeoProf's execution and reads out the profiled statistics periodically via sending commands (\btwo) through an MMIO (Memory-Mapped I/O) interface. In the OS kernel space, we implement the  driver (\bfour) to interact with NeoProf hardware.
 Since NeoProf does not require any modifications to the host CPU, \uline{it is drop-in compatible with any CXL-enabled server platforms.} 

Note that cold pages in fast memory tiers should also be detected and demoted to slow memory to create space for hot page promotion. Since the detection of cold pages does not need a high resolution, as highlighted in~\cite{tmts}, NeoMem  employs the well-established LRU 2Q mechanism\cite{lru2q} in the Linux kernel for the detection of cold pages (\bsix).




\noindent$\bullet$ \emph{State Monitoring.} In a tiered memory system, monitoring runtime states, such as bandwidth utilization, read/write ratios, and page access frequency distribution, is crucial for effective memory tiering. For example, when the slow CXL memory experiences increased bandwidth usage, it becomes advantageous to migrate more hot pages to fast memory. Additionally, in cases where CXL-side memory employs devices with asymmetrical read/write bandwidths, the migration scheduler should better consider their distinct performance characteristics~\cite{sha2023vtmm}. Furthermore, the distribution of page access frequency also counts in hot page classification~\cite{lee2023memtis}. 
Therefore, we incorporate state monitors into NeoProf. The OS retrieves these critical runtime states through NeoProf commands.


\noindent\textbf{NeoMem Daemon.} Within the OS kernel, we implement a NeoMem daemon (\bfive) that interfaces with NeoProf. This daemon is responsible for managing hot page promotions, adhering to a migration policy specified in user space (\beight). Hot pages are migrated by invoking the kernel's page migration functions (\bseven), at intervals set by the \texttt{migration\_interval} parameter. Additionally, the daemon resets NeoProf's states at regular intervals, as determined by the \texttt{clear\_interval} setting.

\noindent\textbf{Migration Policy.} Based on the rich information provided by NeoProf,  NeoMem's migration policy establishes the guidelines for the NeoMem daemon to orchestrate memory profiling and migration (\bseven). This policy decision is made in the user space, enabling customization and tuning by the user. We will introduce the implementation of NeoMem daemon and migration policy in Section~\ref{sec:neomem_software}.

\section{NeoProf Details}
\label{sec:neoprof}




 \begin{figure} [t]
    \centering
    \includegraphics[width=1.0\linewidth]{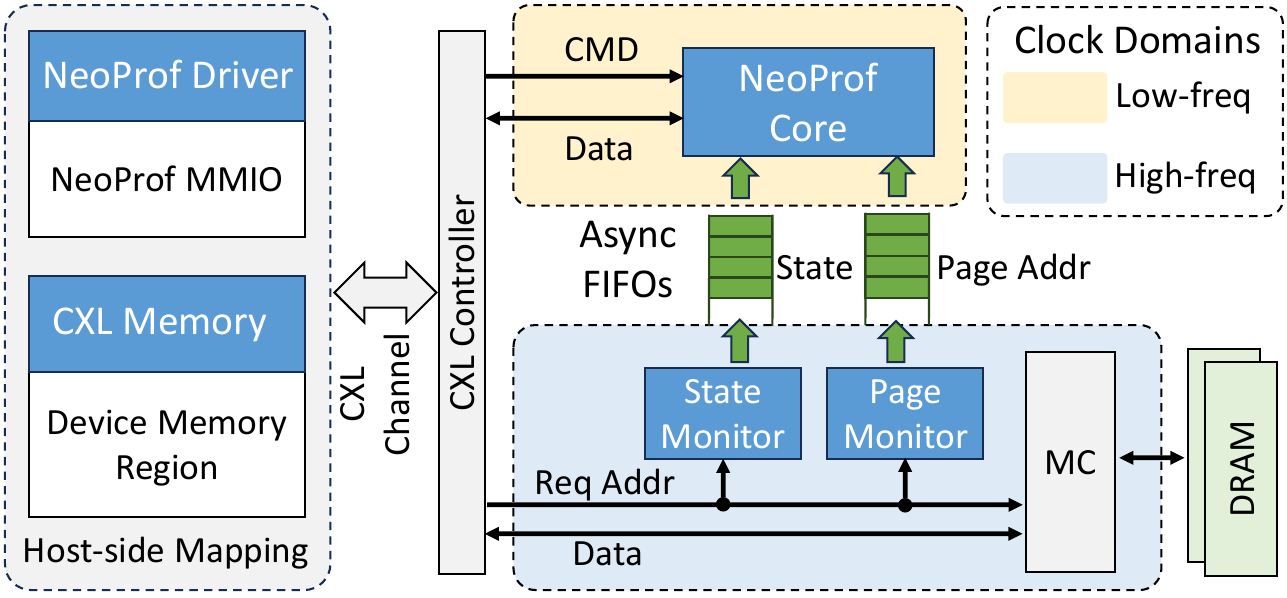} 
  \caption{The Block Diagram of NeoProf.  }
         \label{fig:neoprof}
\end{figure}

\subsection{Architecture Overview}
Figure~\ref{fig:neoprof} illustrates the block diagram of NeoProf hardware, consisting of three custom units: the {State Monitor}, {Page Monitor}, and {NeoProf Core}. The {Page Monitor} snoops  requests from device-side CXL controller, identifies physical page addresses, and forks them to the {NeoProf Core} for analysis. The {State Monitor} tracks Read/Write transactions, estimates bandwidth utilization and read/write ratios, while the {NeoProf Core} collaborates with both monitors, identifying hot pages and responding to host CPU control commands.

As shown in the left part, the registers of NeoProf core are memory-mapped to a predefined address space, enabling the OS kernel to access them via Memory-Mapped IO (MMIO). The host CPU  sends commands to NeoProf by writing  data to specific addresses with different offsets. Additionally, the CXL memory is mapped to another address space, allowing the OS to manage it as a CPU-less NUMA node~\cite{tpp, pond}. 

Notably, current NeoProf prototype is implemented on an FPGA platform. As a result, we place the NeoProf Core in a low-frequency domain, while the State Monitor and Page Monitor operate at a high frequency to match the memory controller's requirements. We use asynchronous FIFOs for data transmission from the monitors to the NeoProf core.

\subsection{Hot Page Detector}


\noindent
\textbf{Challenges.} 
Given a physical page access stream represented as $\mathbb{S} = \{P_{1}, P_{2}, \ldots, P_{n}\}$, where $P_{i}$ denotes the $i$-th accessed page, we define a page as ``hot" if its access frequency surpasses a certain threshold $\theta$.
A straw-man approach for hardware-based hot page detection involves utilizing counters to monitor the access frequency of individual physical pages. However, consider a 512GB CXL memory expander~\cite{samsung_cxl_mem}, housing a total of 128 million 4KB pages. Assigning a 32-bit counter to each page would require 512MB of buffer to store counters. Moreover, updating these counters with every page access would impose a significant burden on DRAM bandwidth. Additionally, reading and processing these counters to identify hot pages could introduce considerable latency, severely affecting the timeliness of profiling.

\begin{figure} [t]
    \centering
    \includegraphics[width=1.0\linewidth]{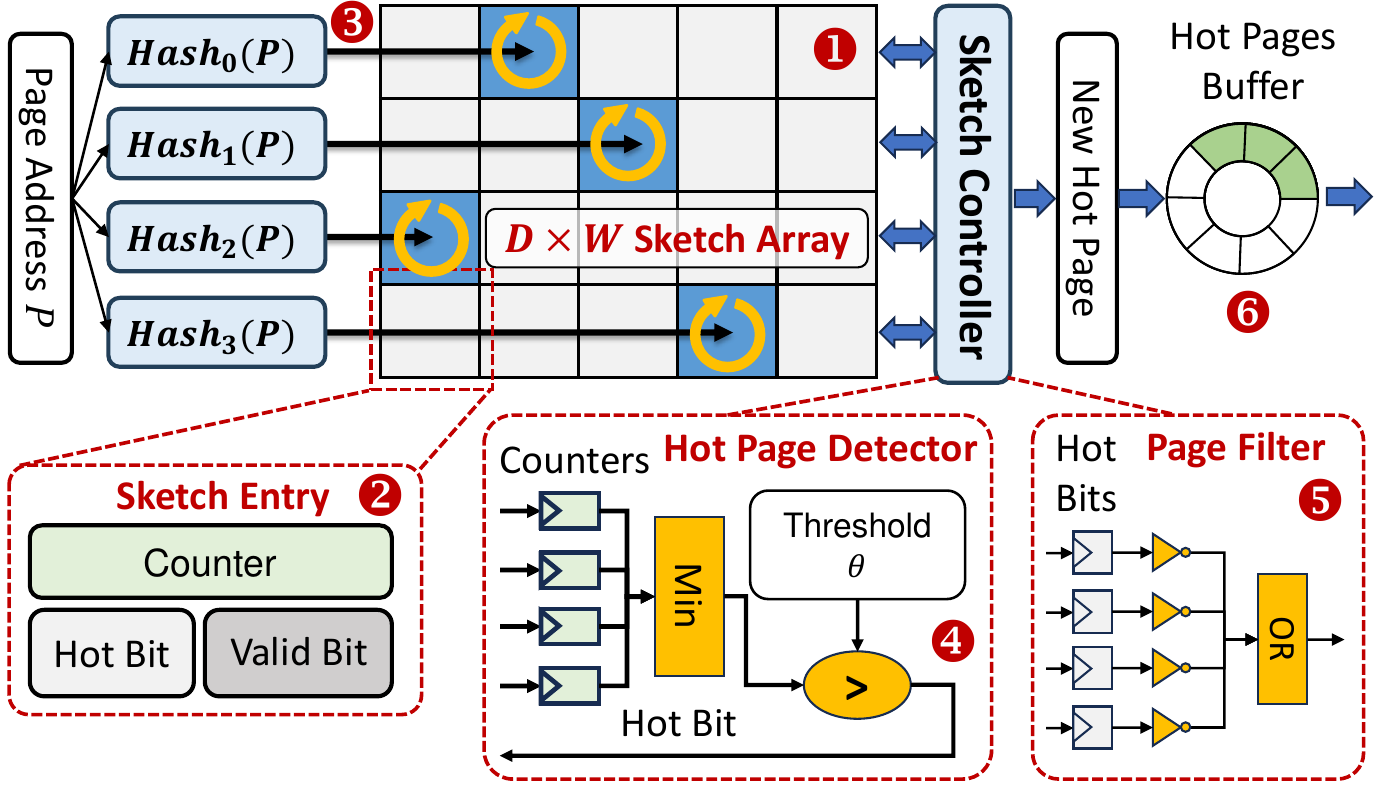} 
    \caption{Hot-Page Detector Architecture.}
         \label{fig:neoprof_details}
\end{figure}

\noindent\textbf{Efficient Hot Page Detection.} To address the challenges outlined, we propose an efficient hot-page detection architecture. Our approach hinges on the Count-Min (CM) Sketch algorithm~\cite{cmsketch}, a hash-based technique  designed for estimating item  frequencies within a data stream.  
We enhance the CM-Sketch with two capabilities: (1) we propose an efficient hot page filtering mechanism to prevent duplication. (2) We introduce an error-bound control mechanism in the hot page detector to guarantee a high detection accuracy.

As illustrated in Figure~\ref{fig:neoprof_details}, a CM-Sketch with parameters ($\epsilon, \delta$) is represented by a two-dimensional array counts (\textcolor{red}{\bone}) with a width of $\mathbf{W}$ and a depth of $\mathbf{D}$. Given parameters ($\epsilon, \delta$), we set $\mathbf{W}= \lceil 2/\epsilon\rceil$ and $\mathbf{D}=\lceil log_2(1/\delta)\rceil$.  Each entry in the array consists of a counter, a hot bit and a valid bit (\textcolor{red}{\btwo}). $\mathbf{D}$ different hash functions (\textcolor{red}{\bthree}) map the input page address to one of $\mathbf{W}$ entries in each lane. 
When a page address $P$ arrives, the hash functions calculate  offsets in each row of the sketching array, denoted as $\Delta_i = h_i(P)$ for the $i$-th row, where $i\in [1, \mathbf{D}]$. Subsequently, the counters of the hashed entries are indexed and incremented:
\begin{equation}
\setlength{\abovedisplayskip}{4pt}
\setlength{\belowdisplayskip}{4pt}
  \Delta_i = h_i(P),~~  A[i][\Delta_i ] \leftarrow A[i][\Delta_i] + 1
\end{equation}
\noindent Then the access frequency of page $P$, represented as $a(P)$, is approximated by the minimum value in each lane: 
\begin{equation}
\setlength{\abovedisplayskip}{4pt}
\setlength{\belowdisplayskip}{4pt}
\hat{a}(P) = \mathop{\mathbf{min}}_{i=1}^{\mathbf{D}} (A[i][\Delta_i]) 
\end{equation}
\noindent According to the theory~\cite{cmsketch},   the estimated access frequency  $\hat{a}(P)$ falls within the following range with probability $1-\delta$: 
\begin{equation}
\label{eq:error_bound}
\setlength{\abovedisplayskip}{4pt}
\setlength{\belowdisplayskip}{4pt}
    a(P) \leq \hat{a}(P) \leq a(P) + \epsilon N
\end{equation}
\noindent Here, $N$ is the total number of accesses seen by the sketch. 
Given a threshold $\theta$, the hot page detector (\textcolor{red}{\bfour}) assesses whether the approximated access count $\hat{a}(P)$ exceeds $\theta$:  
\begin{equation}
\texttt{isHotPage}(P, \theta) = \left\{
\begin{array}{ll}
\texttt{True},            & { \hat{a}(P)  > \theta}\\
\texttt{False},           & { \hat{a}(P)  \leq  \theta}\\
\end{array} \right.
\end{equation}

 NeoProf clears these counters after each hot page detection period, which is implemented via resetting each entry's
  \texttt{Valid} bit. 
 In each sketch lane, the \texttt{Valid} bits are physically arranged in a contiguous manner, allowing for rapid resetting.


\noindent\textbf{Hot-Page Filtering.} In NeoProf's design, the addresses of detected hot pages are put into the hot page buffer (\textcolor{red}{\bsix}~in Figure~\ref{fig:neoprof_details}). However, in each detection period, a page can be identified as hot repeatedly once its access frequency exceeds the threshold $\theta$, which will fill up the hot-page buffer quickly and is redundant for hot page migration. We avoid this problem via introducing a \texttt{Hot} bit in each sketch entry (\textcolor{red}{\btwo}). 
 
Before transferring a page address to the output buffer, the hot page filter (\textcolor{red}{\bfive}) examines  the  \texttt{Hot} bits in the hashed entries.  If all the hot bits are set to \texttt{True}, this suggests that the page might have been recorded previously, leading us to dismiss it. 
Conversely, if any \texttt{Hot} bit is \texttt{False}, this indicates a newly detected hot page.  We then set the hot bits in the corresponding entries to \texttt{True}.
Such a design can be thought of as equivalent to adding a bloom filter~\cite{bloomfilter} after the CM-Sketch unit to probabilistically determine the presence of a hot page~\cite{jin2017netcache}. Our design is more efficient as it reuses the hashing results and introduces only a minimal number of additional hot bits.

\label{sec:error_control}
\begin{figure} [t]
    \centering
    \includegraphics[width=0.98\linewidth]{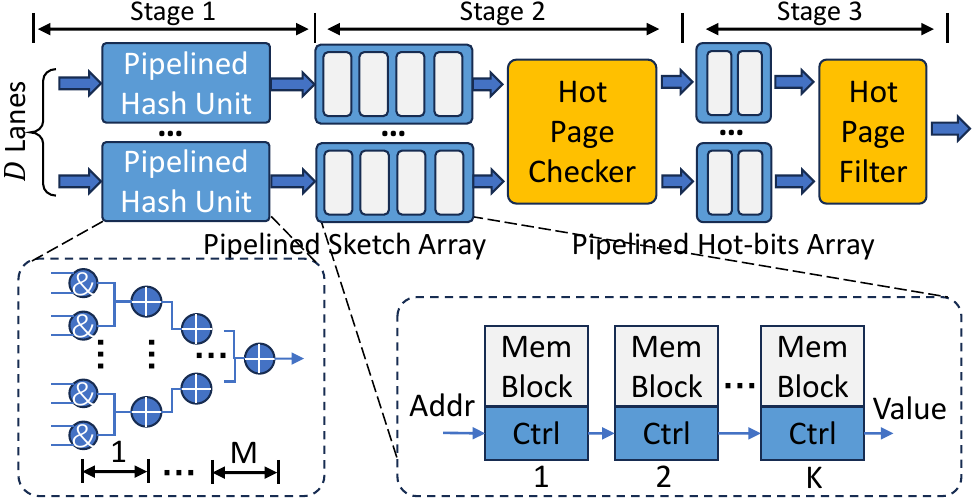} 
    \caption{Hot Page Detector Pipeline.}
         \label{fig:neoprof_pipeline}
\end{figure}

\noindent\textbf{Accurate Error-Bound Estimation.} One significant challenge in Sketch-based hotness estimation is the increasing approximation error as the number of streamed-in page addresses grows ($N$ in Eq.~\ref{eq:error_bound}). In extreme cases, all counters in the sketch array exceeding the threshold $\theta$ can lead to an unseen page being incorrectly labeled as ``hot". Equation~\ref{eq:error_bound} provides a worst-case error bound estimation, which has been criticized as overly ``loose" for practical use~\cite{chen2021precise}.
Chen et al.~\cite{chen2021precise} introduced a technique to estimate a ``near-optimal" approximation error by sorting counters within any row of the sketch array in descending order, denoted as $\{A[1][1], A[1][2], \ldots, A[1][\mathbf{W}]\}$. The tight error bound, referred to as $\mathbf{e}$, is then determined as the $(\mathbf{W} \cdot \lceil\delta^{1/\mathbf{D}}\rceil)$-percentile value of these sorted counters.


With probability $1-\delta$, we can say that $\hat{a}(P) \leq \mathbf{e} + a(P)$. {If} the approximated page access count, $\hat{a}(P)$, exceed threshold $\theta$, we can assert  that $a(P) > \theta - \mathbf{e}$.
For example, given $\mathbf{D} = 2$ and $\delta=0.25$, we can choose the median value of a row to serve as the error estimation, denoted as $\mathbf{e}$. When this error $\mathbf{e}$ comes to a large value (e.g., when $\mathbf{e} > \theta$),  the sketch array may have reached saturation.
NeoMem relies on the estimated error bound to ensure accurate hot page detection.

\noindent\textbf{Hardware Implementation.} Based on these  algorithms, we introduce the hardware implementation details of the proposed hot-page detector and the error-bound estimation logic:
 
\begin{figure} [t]
    \centering
    \includegraphics[width=0.96\linewidth]{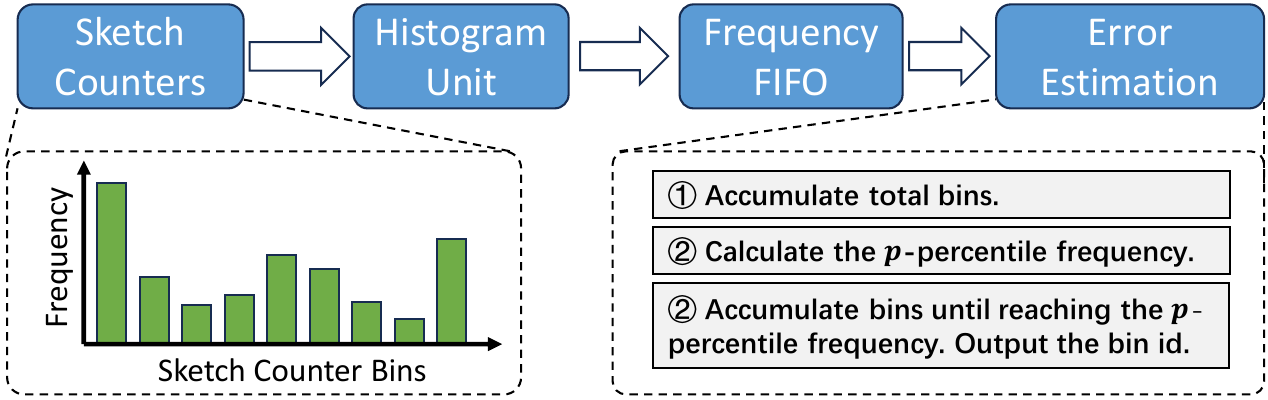} 
    \caption{Histogram-based Error-Bound Estimation}
         \label{fig:error_estimation}
\end{figure}

\noindent$\bullet$ \emph{Pipelined Hot Page Detection.}
In Figure~\ref{fig:neoprof_pipeline}, we break down the detector's pipeline into three primary stages: (1) hash index computation, (2) hot page checking, and (3) hot page filtering. Each  stage comprises finer pipeline stages. We utilize the \emph{H3}~\cite{h3} hash function for hash index computation. This function calculates an $m$-bit hash value  based on a $n$-bit input value $x$   and a $n\cdot m$-bit seed $\pi$: 
\begin{equation}
    h_\pi(x) = x(0)\cdot \pi(0) \oplus x(1) \cdot \pi(1)... \oplus x(n-1)\cdot \pi(n-1)
\end{equation}
where the input $x$ is the $n$-bit page address. $x(i)\cdot \pi(i)$ performs logic \texttt{AND} between each bit $x(i)$ and every bit of $\pi(i)$.
The $\oplus$ operator performs logic \texttt{XOR} operation on vectors. The resulting vector $h_\pi(x)$ is an $m$-bit hashed index. To ensure efficient and pipelined processing, we divide this reduction tree into $\mathbf{M}$ stages. A total of $\mathbf{D}$ pipeline units  handle the hash functions for each sketch row in parallel.

Another challenge arises when the  sketch array width  $\mathbf{W}$ increases. This makes it difficult to achieve a single-cycle index and update for any counter. To address this, we follow prior practices~\cite{tong2017sketch} and  partition the memory into $\mathbf{K}$ sub-blocks and implement the sketch array in a pipelined manner. 

\noindent$\bullet$ \emph{Histogram-based Error-Bound Estimation.} According to the accurate error-bound estimation algorithm introduced above, we should read out a row of sketch counters, sort them, and select the $p$-percentile value as the error bound. To reduce CXL  channel occupation and save host CPU cycles, we propose a histogram-based error-bound estimation mechanism. 

As depicted in Figure~\ref{fig:error_estimation}, NeoProf Core incorporates a histogram unit with 64 bins. Triggered by specific NeoProf commands, the histogram unit  reads the counters in the first row of sketch array and estimates the frequency distribution.   The host CPU just needs to  read out the histogram and estimate the $p$-percentile frequency using a straightforward algorithm. This approach greatly reduces the overheads compared to naively reading out and sorting the entire sketch rows.

Besides facilitating error-bound estimation, \uline{the histogram also approximates the page access frequency distribution in CXL memory.} In Section~\ref{sec:policy}, we will demonstrate how the NeoMem migration policy conducts dynamic hotness threshold adjustment according to the histogram information. 



\begin{table}[t]
\caption{NeoProf commands}
\vspace{-0.2em}
\label{tab:commands}
\resizebox{0.48\textwidth}{!}{
\begin{tabular}{|l|l|l|l|}
\hline
\textbf{Command}         & \textbf{Offset} & \textbf{Operation} & \textbf{Description}                                  \\ \hline
\texttt{Reset}            & 0x100  & Write 1   & Reset NeoProf                           \\ 
\hline
\texttt{SetThreshold}     & 0x200  & Write $\theta$   & Set hot page threshold to $\theta$             \\ 
\texttt{GetNrHotPage} & 0x300  & Read      & Readout \# profiled hotpages          \\ 
\texttt{GetHotPage}      & 0x400  & Read      & Readout a hot page address \\ \hline
\texttt{GetNrSample}      & 0x500  & Read      & Readout \#  sampled cycles \\  
\texttt{GetRdCnt}      & 0x600  & Read      & Readout \# sampled read \\  
\texttt{GetWrCnt}      & 0x700  & Read      & Readout \# sampled write \\ \hline 
\texttt{SetHistEn}      & 0x800  & Write $1$      & Trigger the histogram calculation \\
\texttt{GetNrHistBin}      & 0x900  & Read      & Readout \# histogram bins \\
\texttt{GetHist}      & 0xA00  & Read      & Readout the histogram bins \\ \hline
\end{tabular}
}
\end{table}

\noindent\textbf{NeoProf Commands.}
NeoProf is controlled  by the host CPU  through a set of commands, which are encoded by varying offsets in NeoProf's MMIO region. Some core commands are listed in Table~\ref{tab:commands}. The \texttt{Reset} command clears all the counters and  buffers within NeoProf. The \texttt{SetThreshold} command is used to adjust the hot page threshold $\theta$. Subsequently, the \texttt{GetNrHotPage} and  \texttt{GetHotPage} commands are employed to retrieve the addresses of  hot pages. 
Additionally, we design the \texttt{GetNrSample},  \texttt{GetRdCnt} and \texttt{GetWrCnt} commands to  retrieve  the total sampled cycles, as well as the breakdown of cycles attributed to read and write operations.
Finally, the  \texttt{Hist}-related commands trigger the histogram calculation and retrieve the histogram data. 
\vspace{0.2em}

\section{NeoMem Software Design}
\label{sec:neomem_software}

\subsection{NeoMem Migration Policy}
\label{sec:policy}
Setting the hotness threshold is a critical challenge in memory tiering. Traditional methods, constrained by limited insight into memory access patterns, often rely on static thresholds for classifying hot pages~\cite{corbet2012autonuma, autotiering,yan2019nimble,tpp}, which are sub-optimal. 
With rich and timely memory access information, NeoMem dynamically adjusts the hotness threshold (described in Algorithm~\ref{algo:threshold}) based on the following statistics:


\noindent$\bullet$ \emph{Access Frequency Distribution.} We utilize the page access frequency distribution to dynamically determine the hotness threshold $\theta$. This involves using NeoProf's histogram of sketch counters as a proxy for actual access frequencies (line 4 in the algorithm). The threshold is determined by setting $\theta$ to the $p$-percentile of this distribution. Specifically, we define $Q_F$ as the histogram's quantile function, where $Q_F(x)=y$ implies that a fraction $x$ of pages have fewer than $y$ accesses. Thus, $\theta$ is set to $Q_F(1-p)$ (outlined in line 16), aligning the threshold with the top-$p$ access frequency.

\begin{algorithm}[t]
\def\mathbi#1{\textbf{\em #1}}
\small
\label{algo:threshold}
\caption{Dynamic Hotness Threshold Adjustment}
\textbf{Input:} Migration Quota $m_{quota}$;  Percentile bounds $p_{min},p_{max}$; Default percentile $p_{init}$; \\


$\triangleright$  $p \leftarrow p_{init};$ \\
\While{dynamic threshold adjustment is enabled}{
  $\mathcal{F} \leftarrow get\_neoprof\_hist()$\;
  $\mathcal{B} \leftarrow get\_bandwidth\_util()$\;
  $\mathcal{P} \leftarrow get\_ping\_pong\_count()$\;
  $\mathcal{E} \leftarrow get\_error\_bound(\mathcal{F})$\;
  $\mathcal{M} \leftarrow get\_migrate\_pages\_count()$\; 

  \eIf{$\mathcal{M} < m_{quota}$}{
   $p \leftarrow p\cdot \frac{(1+\mathcal{B})^\alpha}{(1+\mathcal{P})^\beta}$\;
   $p \leftarrow bound(p_{min}, p_{max}, p)$
   }{
   $p \leftarrow max(p_{min}, \frac{p}{2})$;  \textcolor{blue}{ /* Migration quota constraint*/}\
  }
  \If{$Q_\mathcal{F}(1-p) < \mathcal{E}$}{ 
      $p \leftarrow max(p_{min}, \frac{p}{2})$; \textcolor{blue}{ /* Error-bound checking */} \
  }
  $\theta=Q_\mathcal{F}(1-p)$, $update\_hotness\_threshold(\theta)$\;
  $\triangleright$ Wait for the next threshold update period\;
}
\end{algorithm}

\noindent$\bullet$ \emph{Bandwidth Utilization.} We aim for maximum utilization of fast memory in the system. Therefore, when we observe heavy use of the slower CXL memory's bandwidth, it prompts the migration of more pages to the fast memory tier. We define bandwidth utilization ($\mathcal{B}$) as the ratio of memory reads and writes to total sampled cycles, given by $\mathcal{B}=\frac{read + write}{total\ cycles}$ (line 5). {$read$ and $write$ represent cycles when the device is transferring read and write data monitored by NeoProf during the last threshold update period,} and $total\ cycles$ stands for the sampled cycles in that period. 
The hotness threshold should be inversely proportional to $\mathcal{B}$, denoted as $\theta \propto  \frac{1}{\mathcal{{B}}}$. 


\noindent$\bullet$ \emph{Ping-Pong Severity.} An improperly low hotness threshold may lead to a situation where infrequently accessed pages are prematurely promoted to fast memory, only to be swiftly demoted back to slower memory, which is referred to as Ping-Pong phenomenon~\cite{tpp}. To measure ping-pong severity, we introduce the $PG\_demoted$ page flag in the Linux kernel, set when a page is demoted and cleared when it's promoted. A page with the $PG\_demoted$ flag set and then promoted again is counted as a ping-pong event. Ping-pong severity is the ratio of ping-pong events to promoted pages in the previous period, calculated as $\mathcal{P}=\frac{\# ping\ pong\ events}{\# promoted\ pages}$ (line 6). The hotness threshold should be  proportional to $\mathcal{P}$, denoted as $\theta \propto {\mathcal{P}}$.



\noindent$\bullet$ \emph{Approximation Error.} To ensure the precision of hot page classification necessitates considering the approximation error of the sketch-based hot page detector (line 7 of the algorithm). We assume that if the estimated error bound $\mathcal{E}$ exceeds the threshold $\theta$, this indicates considerable inaccuracies in hot page detection. To mitigate this, we  increase the threshold by halving the percentile $p$  (as outlined in lines 14 and 15) to enhance the confidence in  hot-page detection.

\noindent$\bullet$ \emph{Migration Quota.} Lastly, to prevent excessive CPU resource and memory bandwidth consumption due to page migration, we establish a migration quota designated as $m_{quota}$. If the number of migrated pages during the previous period exceed this set quota ($m_{quota}$), we also halve parameter $p$ to make the threshold $\theta$ higher, as outlined in line 13.

Taking these factors into consideration, we calibrate the parameter $p$ to $p \cdot \frac{(1+\mathcal{B})^\alpha}{(1+\mathcal{P})^\beta}$ (line 10), where $\alpha$ and $\beta$ are adjustable hyper-parameters. 
As described in the algorithm, in every threshold update period, we dynamically choose  the top $p$ fraction of pages as hot and make sure that the decision meets the constraints from  error bound and migration quota.










\subsection{User-Space Interface}
In order to facilitate the configuration of runtime parameters in NeoMem, we introduce a set of user-space interfaces, which are accessible through the \texttt{/sys/kernel/mm/neomem} directory. These interfaces are linked to various functions  implemented in the kernel space, which empower users to retrieve essential information from NeoProf and adjust parameters, such as the hotness threshold, migration interval, etc. The migration policy is then implemented within the user space, utilizing these interfaces as its communication channel. Users also have the flexibility to implement their own custom scheduling policies via these interfaces.


\section{Evaluation}
\label{sec:eval}


\subsection{Experimental Setup} 
\label{sec:exp_setup}

\noindent\textbf{Prototyping Platform.} We evaluate NeoMem's practicality and performance on a FPGA-based CXL memory system, detailed in Table \ref{table:system_config}. The setup includes a single-socket Intel$^\circledR$ Sapphire-Rapids$^\texttt{TM}$ CPU and a CXL-enabled Intel$^\circledR$ Agilex$^\texttt{TM}$-7 I-Series FPGA acting as CXL memory (CXL 1.1, Type-3 device). {Intel enables \texttt{cxl.mem} on this FPGA where the CXL-and memory-related IP cores are implemented on the chiplet\cite{cxl_ip}}. The FPGA has dual-channel DDR4-2666 memory with 16GB capacity. The host CPU is equipped with 32GB $\times 4$  DDR5-4800 memory.
For different fast-slow memory ratios, we adjust host memory size by reserving a specific amount of physical memory within the Linux kernel~\cite{memmap}. The default fast-slow ratio is 1:2. We disable CPU's SMT, fix the CPU clock frequency and clear the page cache before running workloads to ensure consistent performance.

\begin{figure} [t]
    \centering
    \includegraphics[width=0.96\linewidth]{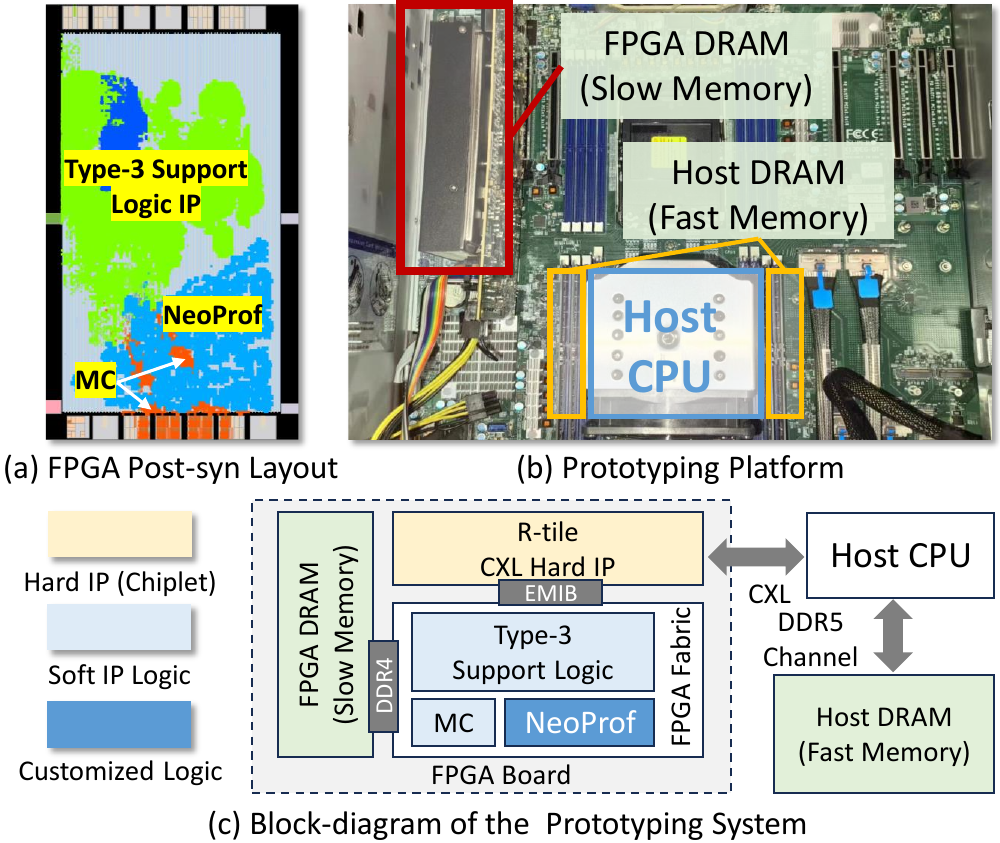} 
    \vspace{-0.5em}
    \caption{{FPGA-based Prototyping System.}}
         \label{fig:exp_platform}
\end{figure}

\begin{table}[t]
\caption{Evaluation System Configuration}
\vspace{-0.5em}
\label{table:system_config}
\resizebox{0.485\textwidth}{!}{

\begin{tabular}{|l|l|}
\hline
\multirow{2}{*}{Host CPU}   & Single socket Intel$^\circledR$ Xeon 6430 CPU @ 2.10GHz                                               \\ \cline{2-2} 
                            & \begin{tabular}[c]{@{}l@{}}32 Cores, hyperthreading disabled. \\ 60MB Shared LLC\end{tabular} \\ \hline
DDR Memory            & 32GB DDR5 4800MHz x 4                                                                          \\ \hline
\multirow{2}{*}{CXL Memory} & {One} Intel$^\circledR$ Agilex$^{\texttt{TM}}$ I-Series FPGA Dev Kit @400 MHz                                                    \\ \cline{2-2} 
                            & \begin{tabular}[c]{@{}l@{}}Hard CXL 1.1 IP on PCIe Gen5 x16\\ 16GB 2-Channel DDR4-2666 DRAM\end{tabular}      \\ \hline
\end{tabular}
}
\end{table}

\begin{table}[t]
\caption{Hardware Parameters of NeoProf}
\vspace{-0.5em}
\label{tab:hardware_params}
\resizebox{0.48\textwidth}{!}{\renewcommand{\arraystretch}{0.6}{
\begin{tabular}{c|c|c|c|c|c}
\toprule
\begin{tabular}[c]{@{}c@{}}Addr \\ Bits\end{tabular} & \begin{tabular}[c]{@{}c@{}}Counter \\ Bits\end{tabular} & \begin{tabular}[c]{@{}c@{}}Sketch\\ Width(\textbf{W})\end{tabular} & \begin{tabular}[c]{@{}c@{}}Sketch\\ Lane(\textbf{D})\end{tabular} & \begin{tabular}[c]{@{}c@{}} \# Memory \\ Segment\end{tabular} & \begin{tabular}[c]{@{}c@{}}Hot Buffer \\ Entries\end{tabular} \\ \midrule
                                           32          &    16                                                     &    512K                                                       &   2                                                       &                128                                          &    16K                                                         \\ \bottomrule
\end{tabular}
}}
\end{table}

\begin{figure*} [t]
    \centering
    \includegraphics[width=1.0\linewidth]{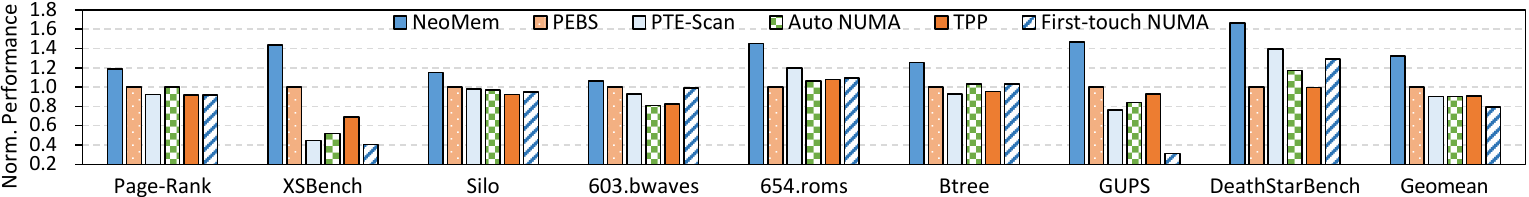} 
    \vspace{-1.8em}
    \caption{End-to-end Performance Comparison.}
    \vspace{-1.2em}
         \label{fig:performance_overview}
\end{figure*}

\begin{figure*} [t]
    \centering
    \includegraphics[width=1.0\linewidth]{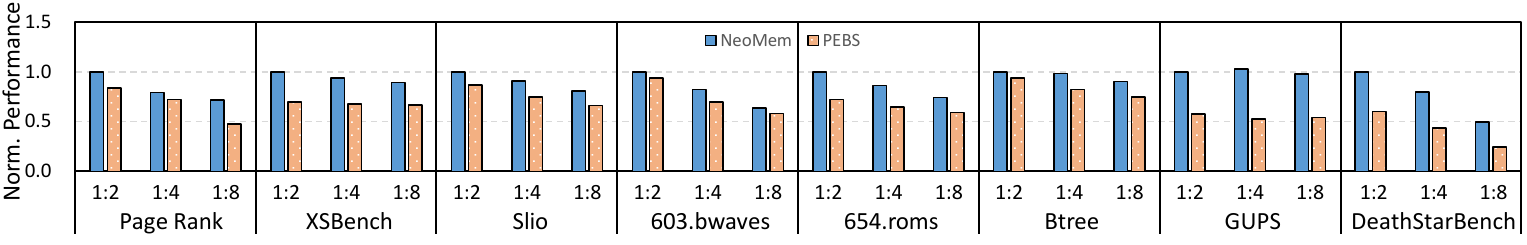} 
    \vspace{-1.8em}
    \caption{{Performance with Different Fast-Slow Memory Ratios.}}
    \vspace{-1.5em}
         \label{fig:different_mem_configs}
\end{figure*}


\noindent\textbf{Benchmarks.} Our evaluation utilizes {eight} representative benchmarks {that have been widely used in previous memory-system studies}: 
DeathstarBench~\cite{gan2019open}, a representative data-center benchmark;
Page-Rank (PR)~\cite{beamer2015gap}, a classic graph processing workload;
XSBench~\cite{Tramm:wy} and GUPS\footnote{The original GUPS has random memory access, we follow HeMem's practice~\cite{raybuck2021hemem} and make some memory access regions hotter than the others.}~\cite{gups}, both are HPC workloads characterized by skewed hot memory regions;  Silo~\cite{tu2013speedy}, an in-memory database for which we employ the YCSB-C workload; Btree~\cite{btree}, an in-memory index lookup workload; and two scientific computing applications from SPEC-2017, namely 603.bwaves and 654.roms, selected for their substantial Resident Set Size (RSS). The RSS values for these benchmarks range from 10.3~GB to 19.7~GB. 
Workloads are executed with 32 threads to fully stress the CPU cores.



\noindent\textbf{Baselines.} We select five baselines for comparative analysis, covering the memory access profiling techniques introduced in Section~\ref{sec:memory_profiling}. Specifically, to compare NeoMem with hint-fault monitoring  methods, TPP~\cite{tpp} and AutoNUMA~\cite{corbet2012autonuma} are chosen. TPP enhances hint-fault monitoring by introducing several new features.
AutoNUMA, part of Linux kernel v6.3,  blends part of TPP's features and introduces configurable hotness threshold.  In addition, to compare with PTE-scan and PMU sampling  methods, we integrate these profiling techniques into NeoMem, replacing its native memory profiling functions. We call these two systems {PTE-scan} and {PEBS} for short.   Lastly, we include First-touch NUMA as a baseline, a widely-used memory allocation policy that assigns pages to the fast memory tier until it's full, without subsequent migration.

\subsection{Implementation}
\label{sec:implementation}

We implement NeoProf hardware in Verilog, {connect it to Intel's Type-3 CXL IP},  and  then synthesize the design using  Quartus 22.3. {Fig.\ref{fig:exp_platform}-c shows the block-diagram of the implemented prototype.} The NeoMem's software parts, along with other baseline methods, are all developed based on  Linux kernel v6.3 for fair comparison.

\noindent\textbf{Hardware Parameters.} Table \ref{tab:hardware_params} lists the default hardware parameters employed by NeoProf.  We configure two sketch lanes ($\mathbf{D}=2$), each equipped with 512K counters ($\mathbf{W}=512K$), where each counter is 16 bits in size. The sketch counter array is divided into 128 pipeline stages. 
We allocate 16K hot page buffers to accommodate detected hot pages. Additionally, we utilize 32 bits to index the device-side page address (4KB page), allowing us to address up to 16TB of memory for each memory controller.

\noindent\textbf{Software Parameters.} The default software parameters of NeoMem and baseline methods are listed in Table~\ref{table:parameters}. For NeoMem, we carefully set the parameters in Algorithm~\ref{algo:threshold}. For baselines, the parameters are also tuned on each benchmark to guarantee a high performance. 

\noindent\textbf{FPGA  Resource Utilization.}  
Our NeoProf implementation mainly consumes 93.8K ALMs (10\%) and  1.5K BRAMs (M20K, 12\%), no DSPs. The FPGA post-synthesize  layout is shown in Figure~\ref{fig:exp_platform}-(a). 
The light blue parts are consumed by NeoProf and the remaining parts are mainly consumed by Intel's FPGA support logic (Type-3 device) for CXL hard IP. 

\begin{table}[t]
\vspace{-1em}
\caption{{Default Software Parameters}}
\vspace{-0.5em}
\label{table:parameters}
\resizebox{0.485\textwidth}{!}{
\setlength{\tabcolsep}{0.8mm}{
\begin{tabular}{l|l|l}
\toprule
{\textbf{Parameter}  }                  & {\textbf{Value} }   & {\textbf{Description} }                     \\ \midrule
{$m_{quota}$}                  &  {$256$MB/s}   & {The maximum page migration rate.} \\ 
{$p_{min}$ }                  & {$0.01\%$} & {The lower percentile bound}       \\ 
{$p_{max}$  }                  & {$1.56\%$} & {The upper percentile bound}       \\ 
{$p_{init}$}                   & {$0.1\%$}  & {The init value of $p$ }           \\ 
{$\alpha$ / $\beta$}          & {$1 / 2$}    & {Adjustable hyper-parameters}      \\ 
{\texttt{migration\_interval}} & {$10ms$}     &   {The interval of  page migration in NeoMem}                         \\ 
{\texttt{clear\_interval}}     & {$5s$}       &   {The interval of  resetting NeoProf
counters}                             \\ 
{\texttt{thr\_update\_interval}}     & {$1s$}       &   {The interval of  updating hot-page threshold}                             \\ 
{\texttt{pebs\_sampling\_rate}} & {$200$-$5000$}    &  {The sampling rate of PEBS}       \\ {\texttt{pte\_sampling\_rate} }& {$1$-$3s$}     & {PTE-sampling rate of TPP\&AutoNUMA}       \\ 
{\texttt{page\_scanning\_rate} }& {$5s$}    & {Page table scanning rate of PTE-scan}       \\ 
\bottomrule
\end{tabular}}
}
\end{table}

\subsection{Main Results}
\label{sec:main_results}
\noindent\textbf{Performance Comparison.} Figure~\ref{fig:performance_overview} shows the performance comparison of our proposed NeoMem system against  baseline systems. All performance numbers are normalized against the {PEBS} system. As depicted by the blue bars representing {NeoMem}, our approach consistently demonstrates superior performance across all seven benchmarks, achieving geomean speedups ranging from {$32\%$} (over {PEBS}) to a remarkable {$67\%$} (over {First-touch NUMA}). {On the representative data-center benchmark, DeathStarBench, NeoMem achieves $1.19\times$ to $1.67\times$ speedup over baseline methods.} These results demonstrate NeoMem's efficiency in tiered memory management.

In certain benchmarks, {NeoMem} exhibits astonishing performance improvements compared to the baseline systems. For instance, {NeoMem} outperforms {First-touch NUMA} by factors of $3.5\times$ and $4.7\times$ in the XSBench and GUPS benchmarks, respectively. {NeoMem} also achieves $2.8\times$ and $3.2\times$ speedup over {AutoNUMA} and {PTE-scan} in XSBench, respectively.  
This remarkable performance gain is attributed to the skewed hot memory regions present in GUPS and XSBench, as discussed in \cite{lee2023memtis}. NeoMem promptly and accurately identifies these hot regions based on NeoProf and efficiently migrates them to the fast memory, thus significantly enhancing system performance. {More detailed analysis of slow-tier traffic reduction of different solutions are presented in Sec.\ref{sec:analysis}}.



\noindent\textbf{Performance with Different Memory Configs.} To illustrate NeoMem's performance under various memory setups, we maintain a constant CXL memory size and investigate three fast-to-slow memory ratios: 1:2, 1:4, and 1:8. Our evaluation compares NeoMem with PEBS, the second-best memory-tiering system according to Figure~\ref{fig:performance_overview}. As depicted in Figure~\ref{fig:different_mem_configs}, NeoMem consistently outperforms PEBS.
Notably, in Page-Rank and Btree, the performance gap between NeoMem and PEBS widens as fast memory shrinks, indicating NeoMem's higher accuracy in hot page classification. Conversely, in GUPS and XSBench, the performance of both NeoMem and PEBS remains relatively stable. This is because the hot  sets  always fit within the fast memory. 


\begin{figure*} [t]
    \centering
    \includegraphics[width=1.0\linewidth]{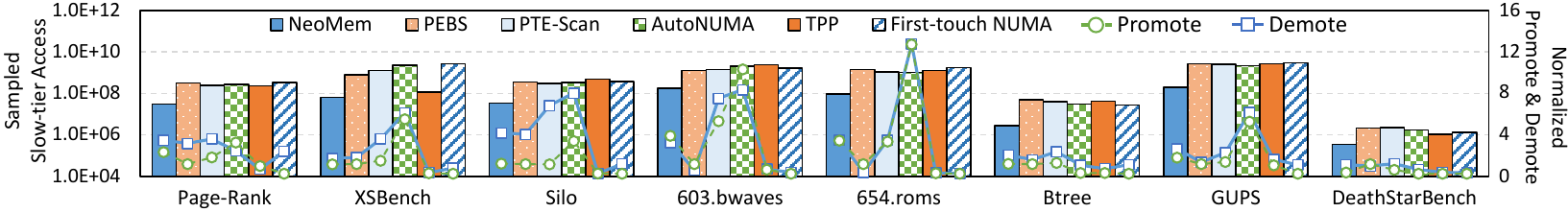} 
    \vspace{-1.2em}
    \caption{{Slow-Tier (CXL Memory) Traffic and \# of Promotions/Demotions Comparison.}}
    \vspace{-1.5em}
         \label{fig:overall_analysis}
\end{figure*}

\begin{figure} [t]
    \centering
    \includegraphics[width=1.0\linewidth]{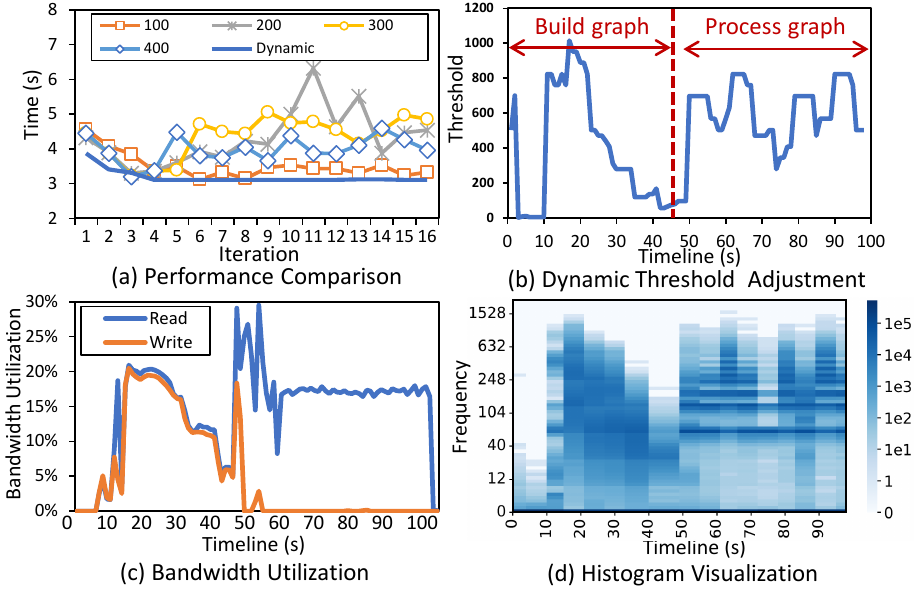} 
    \vspace{-1.2em}
    \caption{Profiling of NeoMem on Page-Rank Benchmark.}
         \label{fig:dynamic_adjustment}
\end{figure}

\subsection{Analysis of NeoMem}

\label{sec:analysis}

\noindent{\textbf{Memory Traffic and Page Migration Analysis.} To better understand per-application behaviors, we profile slow-tier (CXL memory) access and page migration using NeoProf's state monitor and  Linux kernel's counters. As shown in Fig.\ref{fig:overall_analysis}, NeoMem exhibits significantly lower slow-tier traffic across all benchmarks, which explains its superior performance. 
Note that the slow-traffic reduction is not strictly proportional to end-to-end performance in some cases. For example, on XSBench, PEBS has higher slow-tier access than TPP but has a better end-to-end performance in Fig.\ref{fig:performance_overview}.  This is due to other  system-level affecting factors, e.g., false page promotion also incurs slow-tier access, which varies among solutions.
}

{From the figure, NeoMem's promotion count (normalized to PEBS) is significantly  lower than AutoNUMA's, and is on par with PTE-scan. This implies NeoMem's superior ability to  identify hot pages accurately and promptly. TPP exhibits the fewest migration counts in most cases, as it promotes pages only after two consecutive hint-faults. First-touch NUMA performs the worst among all baseline solutions, primarily due to its absence of promotion.  PEBS demonstrates fewer promotions than NeoMem in the majority of cases, this suggests that its sampling-based tracking has low coverage and is prone to missing a large number of hot pages.
}

\noindent{\textbf{CPU Overhead of NeoMem.} As NeoMem offloads memory profiling to dedicated hardware, the host CPU only has to retrieve data from NeoProf, which has minimal overhead. To prove this, we evaluate the slowdown on the GUPS benchmark relative to a baseline system where NeoProf is disabled. After several trials, we observe a mere 0.021\% slowdown.}

\noindent\textbf{Effectiveness of the Migration Policy.} 
{To demonstrate the effectiveness of our NeoMem policy introduced in Section~\ref{sec:policy}, we compare it to the naive fixed-threshold policy.}
We consider the Page-Rank workload processing a graph through sixteen iterations. In each iteration, the execution time is recorded. We compare NeoMem's dynamic threshold policy with fixed thresholds ($\theta = \{100, 200, 300, 400\}$). 

As depicted in Figure~\ref{fig:dynamic_adjustment}-(a),  the dynamic threshold policy employed by NeoMem (the dark blue line) consistently results in the shortest execution times across these iterations. Fixed thresholds, for example, $\theta=200$, suffer from an obvious slowdown after the $9$-th iteration. This outcome demonstrates the effectiveness and necessity of a dynamic scheduling policy.

Figure~\ref{fig:dynamic_adjustment}-(b) illustrates the evolving hotness threshold during Page-Rank testing. The threshold is  dynamically adjusted from 1 to 1000, with initial low settings and rapid increases (50s to 100s) in response to runtime conditions. 

In Figure~\ref{fig:dynamic_adjustment}-(c), we plot the runtime read/write bandwidth profiled by NeoProf. We can observe that during the initial graph processing phase (around 40s), high bandwidth utilization prompts NeoMem to set a low threshold (as seen in Figure~\ref{fig:dynamic_adjustment}-(b)), promoting more pages and effectively reducing CXL memory bandwidth utilization.

Figure~\ref{fig:dynamic_adjustment}-(d) visualizes the evolving access frequency histogram profiled by NeoProf. Every 5 seconds, the profiled histogram is plotted as the vertical strip. The darker regions  represent that more pages have this page-access frequency. The distribution of dark regions appears to correspond closely with the fluctuations observed in Figure~\ref{fig:dynamic_adjustment}-(b).  This suggests that  hotness threshold is properly set according to the runtime page access frequency distribution.

\begin{figure} [t]
    \centering
    \includegraphics[width=0.98\linewidth]{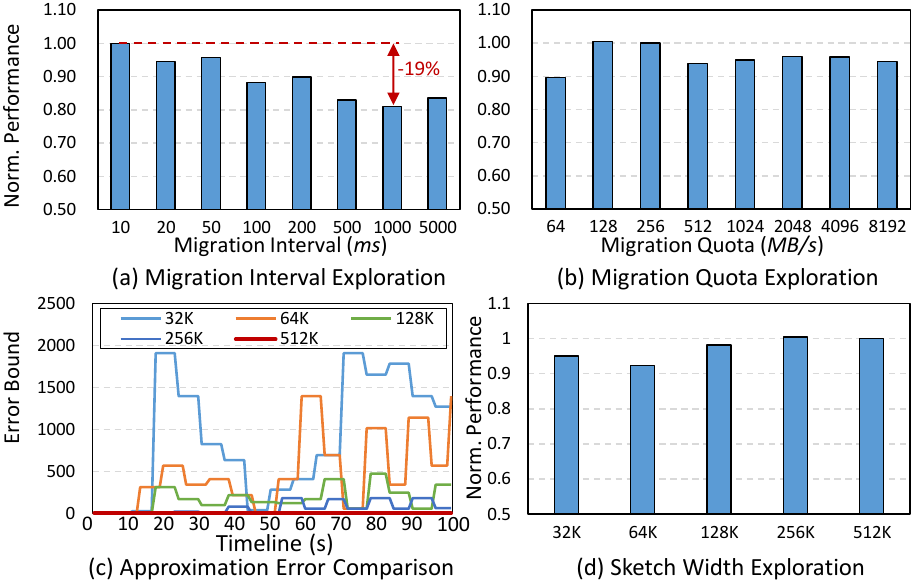} 
        \vspace{-0.1em}
    \caption{Sensitivity to System and NeoProf Parameters.}
         \label{fig:sensitivity}
\end{figure}



\noindent\textbf{Sensitivity to System Parameters.} We investigate the impact of \texttt{migration\_interval} (the period at which NeoMem retrieves hot pages from NeoProf and does promotion) and migration quota ($m\_quota$ in Sec. \ref{sec:policy}) on system performance.

\noindent$\bullet$ \emph{Migration Interval.} In Figure~\ref{fig:sensitivity}-(a), we vary the migration interval from 10ms to 5000ms and assess its effect on the Page-Rank benchmark. A shorter migration interval generally results in better performance, as it enables more timely detection and migration of hot pages. Achieving a short migration interval requires a memory profiling technique with high time resolution and low overhead, highlighting NeoProf's advantages. In comparison, PTE-scan based methods can only support second-level hot page detection and migration~\cite{raybuck2021hemem, amp}.

\noindent$\bullet$ \emph{Migration Quota.} In Figure~\ref{fig:sensitivity}-(b), we vary the migration quota from 64MB/s to 8192MB/s and evaluate its impact on performance. We find that a 64MB/s migration quota results in a 10\% lower performance compared to 128MB/s or 256MB/s.  Increasing the migration quota further slightly hampers overall performance due to heightened migration aggressiveness.


\noindent \textbf{Sensitivity to NeoProf  Parameters}.We evaluate NeoMem's sensitivity to NeoProf's hardware parameters, specifically sketch width ($\mathbf{W}$) and sketch lanes ($\mathbf{D}$). We find that using a single lane ($\mathbf{D}$=$1$) results in a  decrease in end-to-end performance. However, increasing $\mathbf{D}$ beyond 2 does not improve performance obviously. 
Consequently, we empirically choose to maintain $\mathbf{D}$ at 2 in our prototype.
We then vary the parameter $\mathbf{W}$ from 32K to 512K and plot both the error-bound curve, calculated using the algorithm in Section~\ref{sec:neoprof}, and the system performance using the Page-Rank benchmark. As shown in  Figure~\ref{fig:sensitivity}-(c), increasing $\mathbf{W}$ dramatically reduces the error-bound, which is constantly zero when the sketch width reaches 512K. Concurrently, system performance improves with the increasing of sketch width, peaking when $\mathbf{W}$= 256K, as shown in Figure~\ref{fig:sensitivity}-(d). Our prototype sets $\mathbf{W}$ at 512K to ensure a sufficiently low error bound and high performance. 



\noindent\textbf{Convergence Analysis on GUPS.} 
{To examine how the low overhead and high resolution/accuracy advantages of NeoProf 
contribute to improved performance, we perform a convergence analysis using the GUPS microbenchmark.}
In this experiment, we confine 90\% of memory access to a fixed memory region, while the remaining 10\% of memory access uniformly falls in the whole working set.  For each method we warm up the system for 600s to reach a convergence. Then we suddenly change the location of the hot set to evaluate the convergence speed of different methods. 

\begin{figure} [t]
    \centering
    \includegraphics[width=0.98\linewidth]{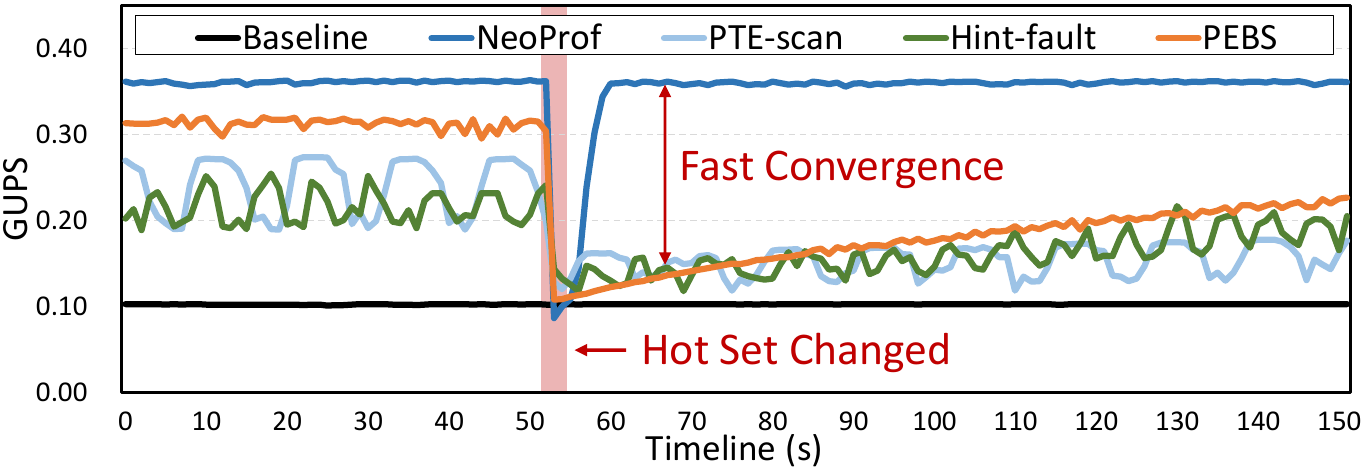} 
    \vspace{-0.5em}
    \caption{Comparison Among Different Profiling Methods.}

    \label{fig:motivation_converge}
\end{figure}

As shown in Figure~\ref{fig:motivation_converge}, we plot the  GUPS (giga updates per second, higher is better) of NeoProf and other baselines over time. \rev{The \texttt{pebs\_sampling\_rate} for PEBS based method is set to 397 in this experiment. }
NeoProf shows the highest GUPS in the converged state (0-50s), indicating that it accurately classifies the hot pages and cold pages, avoiding unnecessary page migration. The smooth curve of NeoProf also reveals that NeoProf has a low overhead. After the hot set change at about 50s, NeoProf shows the fastest converge speed, indicating that NeoProf quickly identifies hot pages and migrates them to local memory, thanks to the high time resolution and  space resolution.

\vspace{1em}

\section{Discussion \& Future Work}
\label{sec:discussion}


\begin{table}[t]

\caption{{\rev{Transparent Huge Page VS. Base Page on Page-Rank}}}
\vspace{-0.5em}
\label{table:huge_page}
\resizebox{0.485\textwidth}{!}{
\setlength{\tabcolsep}{0.8mm}{
\begin{tabular}{l|c|c|c|c}
\toprule
        \rev{{Memory-Tiering Technique}}          & \rev{NeoMem THP} & \rev{TPP THP} & \rev{NeoMem Base} & \rev{TPP Base} \\ \midrule
 
 {Generate (s)} & \rev{7.61}      & {10.96}    & {8.63}        & \rev{10.67}    \\ 
  {Build  (s)} & \rev{23.90}      & \rev{36.90}    & \rev{25.58}        & \rev{35.67}     \\ 
 \rev{Avg. Trail (s)} & \rev{2.80}      & \rev{3.59}    & \rev{2.95}        & \rev{3.32}     \\ 
\rev{Total Time (s) }   &  \rev{76.28}          &  \rev{105.31}        &      \rev{81.39}     &  \rev{99.39}       \\ 
\rev{Promoted Base Pages (GB)}&      \rev{11.53}   &  \rev{2.70}       &    \rev{ 14.90}             &  \rev{2.01}   \\ 
\rev{Promoted Huge Pages (GB)} &    \rev{7.02}        &    \rev{0.74 }    &      \rev{/}       &  \rev{  /}      \\ \bottomrule
\end{tabular}
}
}
\end{table}

\noindent\rev{\textbf{Huge Page Support.}
In the experiments discussed earlier, NeoMem and all baselines manage and migrate pages at the  base page (4KB) level. However, NeoMem can also support huge pages since it utilizes Linux's huge-page-compatible page migration functions. NeoProf still reports hot 4KB pages, and the host can migrate huge pages, provided the profiled hot 4KB pages are part of huge pages. To evaluate the performance, we enable the widely adopted Transparent Huge Page (THP) feature in the Linux kernel, enabling the automatic consolidation of base pages into larger, 2MB huge pages. As detailed in Table~\ref{table:huge_page}, NeoMem, when equipped with THP, demonstrates superior performance over the base-page-only configuration on Page-Rank. It efficiently migrates 7.02GB of huge pages into faster memory. In contrast, TPP experiences a  performance decline with THP enabled, migrating only a minimal amount of huge pages. This is mainly due to its low  time-resolution for hot page detection.}

\rev{NeoMem is also orthogonal to earlier memory tiering methods optimized for huge pages, which detect hot pages  and migrate them at base-page granularity to improve  resolution and reduce overhead~\cite{agarwal2017thermostat, lee2023memtis}. }

\begin{figure} [t]
    \centering
    \includegraphics[width=0.98\linewidth]{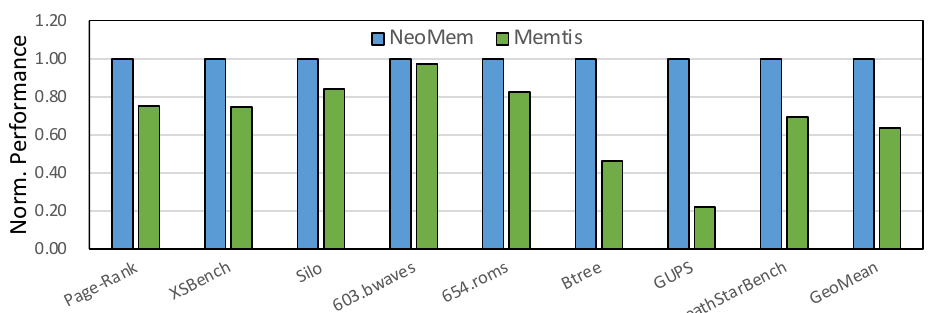} 
    \caption{{ \rev{End-to-end Comparison with Memtis}}.}
         \label{fig:memtis}
\end{figure}

\noindent\rev{\textbf{End-to-end Comparison with Memtis.} Memtis~\cite{lee2023memtis} is a recent  CXL/NVM memory tiering solution. Memtis adopts PEBS to profile memory access, and incorporates dynamic hot set classification based on memory access distribution. We port their released code\footnote{\url{https://github.com/cosmoss-jigu/memtis}} to our hardware platform (with some bugs fixed) and perform end-to-end comparison with NeoMem. 
As shown in Fig.\ref{fig:memtis}, Memtis closely matches NeoMem's performance on 603.bwaves but significantly underperforms on GUPS. NeoMem outperforms Memtis by a $1.58\times$
 geomean speedup. Analysis reveals that Memtis promotes only 1.1\% of pages compared to NeoMem, likely due to its limited ability to adapt to rapidly changing memory access patterns through PEBS and histogram-based hot-page classification.  }

\begin{figure} [t]
    \centering
    \includegraphics[width=0.98\linewidth]{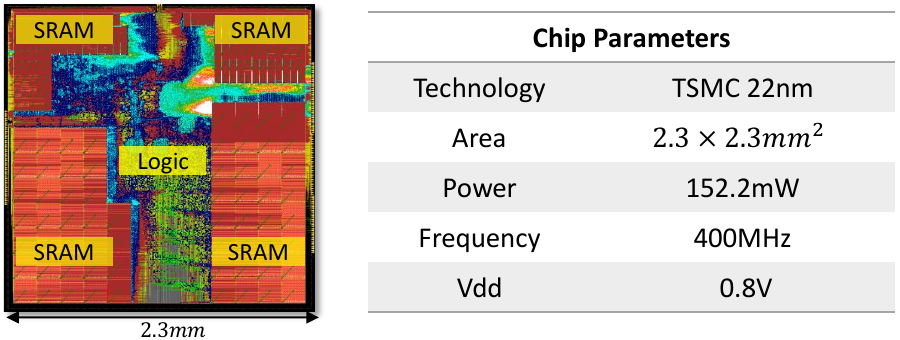} 
    \vspace{-0.5em}
    \caption{{ Layout and Parameters of NeoProf Under TSMC 22nm Process}.}
         \label{fig:asic}
\end{figure}

\noindent\textbf{{Hardware Overhead Estimation.}} {To estimate the hardware cost of integrating NeoProf into  CXL controllers},  we also evaluate the area and power overheads of NeoProf  {using EDA tools}. Figure~\ref{fig:asic} illustrates the layout and parameters of {NeoProf}. This implementation utilizes the TSMC 22nm technology node, with sketch parameters set at $\mathbf{W}$=$256\text{K}$, $\mathbf{D}$=$2$.
The resulting design occupies an area of approximately 5.3$\,mm^2$ and consumes 152.2$\,mW$ of power, which is lightweight  for integration into device-side controllers. \rev{The floor-planing results demonstrate that the SRAM macros occupy about 54\% of chip area, which are used to implement the sketch array, hot-bit array and hot-page buffer. The remaining on-chip area is consumed by NeoProf's compute and control logic.}


\noindent\textbf{Virtualization Support.} In cloud environments based on virtual machines, NeoMem can be integrated into the host OS. The host OS identifies hot physical pages through the NeoMem daemon and executes hot page promotion. Following hot page migration, the Enhanced Page Table (EPT) of the guest virtual machines will undergo remapping~\cite{sha2023vtmm}. Evaluation in virtualized environments is planned for our future work.


\noindent\textbf{Scalability of NeoMem.} In our current prototyping system, we are constrained by a single 16GB CXL memory device due to hardware limitations. However, 
\rev{unlike the PTE-scan baseline, which experiences a linear increase in profiling overhead with memory size due to scanning all pages, NeoProf maintains consistent performance regardless of memory size. This is because NeoProf directly tracks CXL.mem requests, and the maximum request rate is constrained by channel bandwidth, not memory size. Additionally, according to sketching theory, profiling accuracy depends on the volume of incoming requests, which is also bandwidth-related, not memory-size-related. Also, the profiling throughput should linearly scale with the addition of more CXL memory devices equipped with NeoProf. Given the negligible software overhead of host-NeoProf interaction in a single-device scenario (0.021\% slowdown reported in previous section), adding multiple devices does not  burden the hosts. We leave the evaluation in multi-device scenarios to our future work. }

\noindent\textbf{Memory Interleaving.}  In multi-device scenarios,  a single physical page  can be interleaved among multiple devices.
Under such circumstances, the NeoProf in each device only profiles a specific fraction of a page. How will interleaving affect the overall memory-tiering performance is yet to be explored.  The host OS may also need to gather fragmented page hotness information from all NeoProfs and conduct additional post-processing tasks like hot-page de-duplication.


\section{Related Work}

\label{sec:related_work}

\subsection{{Probabilistic Algorithms for Data Stream Analysis}}
\label{sec:sketching_works}
{ Probabilistic algorithms have been extensively employed to solve various tasks such as identifying the presence of specific items in data streams, exemplified by the Bloom Filters~\cite{bloomfilter, countingbloomfilter}. They are also used to identify unique elements, as demonstrated by the HyperLogLog algorithm~\cite{flajolet2007hyperloglog}, and to estimate item frequencies, as shown by the Sketch algorithms~\cite{cmsketch, yang2017pyramid, goyal2011lossy, yang2018elastic, li2022stingy}. In this study, we treat hot page detection as a problem of identifying ``heavy hitters" in memory access streams, a task for which the Count-Min Sketch  algorithms are particularly well-suited~\cite{cmsketch}. 
 }
\subsection{Software-based Tiered Memory System}

Software-based tiered memory systems have seen extensive research in techniques related to page access profiling~\cite{sha2023vtmm, amp, choi2021dancing, tmts, agarwal2017thermostat, refault}, page classification~\cite{amp, lee2023memtis, raybuck2021hemem, ebm, multi-clock }, and efficient page migration~\cite{yan2019nimble, ryoo2018case}. These methods, however, are hindered by limited memory access profiling capabilities, as demonstrated in our work. An alternative approach involves managing memory objects directly at the application~\cite{hildebrand2020autotm, li2022GCMove, ren2021sentinel, Wang2019Panthera, memkind, pmdk} or library~\cite{SMT, dulloor2016data, oh2022MaPHeA, wu2017unimem, wei2015_2pp} level, but these require modifications to users' applications or libraries. Our NeoMem solution offers a practical resolution to these challenges.

\subsection{Architecture Support for Memory Tiering} 
Besides software-based approaches, previous works have also optimized heterogeneous memory systems from a pure architectural standpoint. MemPod~\cite{mempod} uses the Majority Element Algorithm (MEA) to identify hot pages but assumes management of both slow and fast memory by the same hardware, which differs from the current CXL memory system. Similar approaches like CAMEO~\cite{chou2014cameo}, PoM~\cite{sim2014transparent}, and SILC-FM~\cite{ryoo2017silc} treat fast memory as a hardware-managed cache for slow memory.
A recent work, HoPP~\cite{li2023hopp}, suggests modifying CPU memory controllers to track memory accesses and provide information to the OS. However, these approaches necessitate costly CPU-side modifications.  NeoMem is a ``CXL-native" solution, limiting hardware modifications to the device side and avoiding expensive CPU-side upgrades. 


\section{Conclusion}
This paper introduces NeoMem, a novel CXL-native memory-tiering technique. NeoMem embodies a hardware-software co-design philosophy, with the integration of a dedicated hardware profiler called NeoProf into the controllers of CXL memory. This enables the OS to access profiled information and execute efficient hot page migration based on a customized migration policy. Comprehensive evaluation on a real CXL memory platform demonstrates that NeoMem achieves a geomean speedup ranging from {32\% to 67\%} across various existing memory tiering solutions.
\section*{Acknowledgments}

We thank all the reviewers from ISCA 2024 and MICRO 2024 for their valuable comments. We thank Yijin Guan for his kind help.  This work is supported by National Natural Science Foundation of China (NSFC) (Grant No. 62032001) and 111 Project (B18001). This work is also supported in part by the NSFC under Grant No. U21B2017.
Dr. Jie Zhang is supported in part by the National Key Research and Development Program of China (Grant No. 2023YFB4502702) and the Natural Science Foundation of China (Grant No. 62332021).


\bibliographystyle{IEEEtranS}
\bibliography{refs}
\end{document}